\documentstyle[aps,multicol,pre]{revtex}

\input epsf
\newcommand{\dir}{.}
\newcommand{\fig}[4]
{
     \noindent
     \unitlength=1mm
     \begin{picture}(#2,#3)
     \put(0,0){\leavevmode \epsfxsize=#2mm \epsffile{\dir/#1}}
     \end{picture}
   \noindent
#4
}
\begin{document}

\title{Surface induced disorder in body-centered-cubic alloys}
\author{F. F. Haas, F. Schmid${}^{\ddag}$, K. Binder}
\address{
Institut f\"ur Physik, Universit\"at Mainz, D-55099 Mainz, Germany \\
${\ddag}$ Max-Planck-Institut f\"ur Polymerforschung, Ackermannweg 10, 
D-55021 Mainz, Germany}

\maketitle
\tighten

\begin{abstract}
We present Monte Carlo simulations of surface induced disordering in a model
of a binary alloy on a bcc lattice which undergoes a first order bulk 
transition from the ordered DO3 phase to the disordered A2 phase.
The data are analyzed in terms of an effective interface Hamiltonian 
for a system with several order parameters in the framework of the linear 
renormalization approach due to Br\'ezin, Halperin and 
Leibler. We show that the model provides a good description of the system in 
the vicinity of the interface. In particular, we recover the logarithmic
divergence of the thickness of the disordered layer as the bulk transition
is approached, we calculate the critical behavior of the maxima of the layer 
susceptibilities, and demonstrate that it is in reasonable agreement with 
the simulation data. Directly at the (110) surface, the theory predicts that 
all order parameters vanish continuously at the surface
with a nonuniversal, but common critical exponent $\beta_1$.
However, we find different exponents $\beta_1$ for the order parameter
$(\psi_2,\psi_3)$ of the DO3 phase and the order parameter $\psi_1$
of the B2 phase. Using the effective interface model, we derive the 
finite size scaling function for the surface order parameter and show
that the theory accounts well for the finite size behavior of 
$(\psi_2,\psi_3)$, but not for that of $\psi_1$. 
The situation is even more complicated in the neighborhood of the
(100) surface, due to the presence of an ordering field which couples
to $\psi_1$.
\end{abstract}


\begin{multicols}{2}

\section{Introduction}

First order phase transitions in the bulk of systems can drive a variety of 
interesting wetting phenomena at their surfaces and interfaces. They
have attracted much attention over many years\cite{wetting}, and are still 
very actively investigated\cite{wetting2}.  
Prominent examples are the wetting of a liquid on a 
solid substrate at liquid-vapour coexistence, or the wetting of one component 
of a binary fluid below the demixing temperature on the walls of a container.
These systems are representatives of a generic situation, which has been 
studied in particular detail: Three phases coexist, substrate, liquid and 
vapour. The substrate acts as inert ``spectator'' which basically provides the
``boundary conditions'' for the liquid-vapour system. The liquid-vapour 
transition can be described by a single order parameter ({\em e.g.}, the 
density), which can take two equilibrium bulk values at coexistence 
(the liquid density or the gas density). 
Obviously, the liquid phase will only wet the substrate 
if it is preferentially adsorbed by the latter.
As one approaches the liquid-vapour coexistence from the vapour side,
different scenarios are possible, depending on the substrate interactions
and on the temperature: Either the liquid film covering the substrate 
remains microscopic at coexistence (``partial wetting''), or it grows 
macroscopically thick (``complete wetting'').
The transition from partial to complete wetting can be first order or
continuous (``critical wetting'').
Since critical wetting is only expected
on certain substrates at a specific temperature, it is rather difficult
to observe experimentally (An experimental observation of critical wetting 
with long range forces has been reported in \cite{ragil}, and with short 
range forces in \cite{ross}).

Wetting phenomena are also present in alloys which undergo a discontinuous 
order-disorder transition in the bulk\cite{lipowsky1,kroll}. 
In many cases, surfaces are neutral with respect to the symmetry of the 
ordered phase, but reduce the degree of ordering due to the reduced number of 
interacting neighbors. 
The surfaces can thus be wetted by a layer of disordered alloy,
{\em i.e.}, ``surface induced disorder'' (SID) occurs\cite{lipowsky2,dosch1}. 
The situation is reminiscent of liquid-vapour wetting; 
however, the underlying symmetry in the system restricts the possible 
wetting scenarios significantly.

We shall illustrate this for systems with purely short range interactions:
We consider a Landau free energy functional of the form
\begin{eqnarray}
\label{landau}
{\cal F}\{ {\bf m} \} & = & 
\int \! d \vec{r} \! \int_0^{\infty} \!\!\!\! dz \; \Big\{ \;
\frac{g}{2} (\nabla {\bf m})^2 + f_b\Big({\bf m}(\vec{r},z)\Big) \; \Big\} 
\nonumber \\
&& \qquad \qquad
+ \int \! d \vec{r} \; f_s \Big({\bf m}(\vec{r},z=0)\Big).
\end{eqnarray}
Here the vector {\bf m} subsumes the relevant order parameters, 
the $z$-axis is taken to be perpendicular to the surface, and $d\vec{r}$ 
integrates over the remaining spatial dimensions. 
The offset of the bulk free energy
density $f_b({\bf m})$ is chosen such that $f_b({\bf m}_b)=0$ in the bulk.
The surface contribution $f_s({\bf m})$ accounts for the influence of the 
surface on the order parameter, {\em i.e.}, the preferential adsorption of 
one phase or in the case of SID the disordering effect.
In mean field approximation, the functional (\ref{landau}) is minimized
by the bulk equation
\begin{equation}
g \; \frac{d^2 m_i}{d z^2} = \partial_i f_b({\bf m}),
\end{equation}
which describes the motion of a classical particle of mass $g$ in the
external potential $(-f_b({\bf m}))$, subject to the boundary condition 
at $z=0$
\begin{equation}
g \; \frac{d m_i}{d z} = - \partial_i f_s({\bf m})
\quad \mbox{with} \quad
\Big| g \frac{d{\bf {\bf m}}}{dz} \Big| = \sqrt{2 g f_b({\bf m})}
\end{equation}
\end{multicols} \twocolumn

\begin{figure}[t]
\noindent
\fig{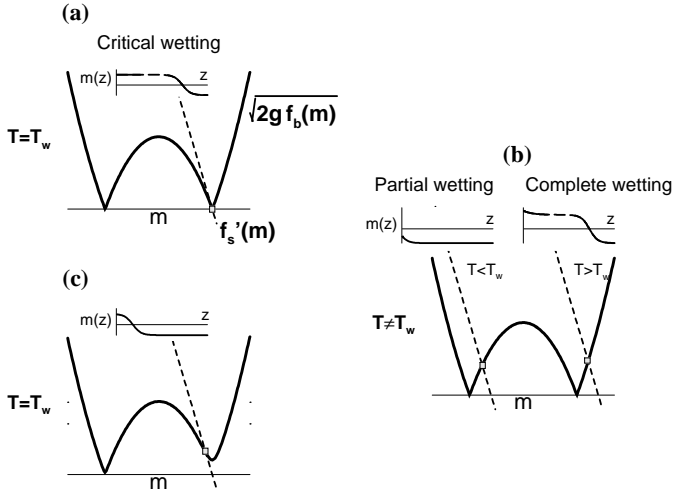}{75}{75}{
\vspace*{-0.4cm}
\caption{ Cahn construction (schematic) for a second order wetting transition:
(a) Critical wetting,
(b) Partial and complete wetting,
(c) Off bulk coexistence, approaching critical wetting.
Insets show the corresponding order parameter profiles.
See text for more explanation.}
\label{cahn_wet}
}
\end{figure}

If the order parameter has just one component, this equation can be 
solved graphically by the Cahn construction\cite{cahn}. 
This is illustrated in Fig. 1 for 
the case of a continuous wetting transition. The corresponding order parameter 
profiles are shown as insets. Complete wetting is encountered if $f_s'(m)$ 
crosses $\sqrt{2 g f_b}$ at the outer side of the minimum corresponding to the 
adsorbed phase. Partial wetting is found if the crossing point is located 
between the two minima (Fig. \ref{cahn_wet}b). Critical wetting connects the 
two regimes, {\em i.e.}, $f_s'(m)$ crosses $\sqrt{2 g f_b}$ right at the 
adsorbed phase minimum of $f_b$ (Fig. \ref{cahn_wet}a). Fig. \ref{cahn_wet}c) 
shows a case where the system is off bulk coexistence. 

Now, let us consider the case of surface induced disorder. Here, several
equivalent ordered phases exist, and the ordered state breaks a symmetry.
For neutral surfaces which do not discriminate between the ordered phases, 
$f_b$ and $f_s$ have the same symmetry. This implies that $f_s$ is extremal 
in the disordered phase (${\bf m}=0$), {\em i.e.}, $|\partial f_s|$ is zero 
at ${\bf m}=0$ and thus crosses $\sqrt{2 g f_b}$ at ${\bf m}=0$ 
\begin{figure}[b]
\noindent
\hspace*{-0.3cm}\fig{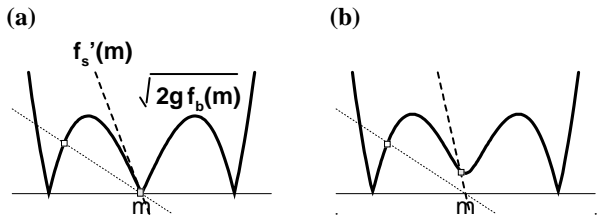}{80}{37}{
\vspace*{-0.3cm}
\caption{Cahn construction (schematic) for surface induced disorder in a 
system with a one-component order parameter $m$ (a) at bulk coexistence and 
(b) off bulk coexistence. 
Dashed line shows surface term $f_s'(m)$ for critical wetting, 
dotted line for partial wetting.}
\label{cahn_sid}
}
\end{figure}\noindent
(Fig. \ref{cahn_sid}). 
Comparing that with the scenario sketched above (Fig. \ref{cahn_wet}), 
we find that surface induced disordering corresponds to 
either partial or critical wetting\cite{kroll} 
-- the symmetry of the surface interactions
excludes the possibility of complete wetting\cite{fn1}.
The off-coexistence situation (Fig. \ref{cahn_sid}b)) resembles
that in Fig. \ref{cahn_wet}c).

Alloys which exhibit surface induced disorder thus seem particularly 
suited to study critical wetting. Unfortunately, the simplification
due to the symmetry of the system often goes along with severe
complications in other respect: Usually, one has to deal with a number
of order parameters and other coupled fields, which interact in a way that 
may not always be transparent. If the surface under consideration does
not have the symmetry of the bulk lattice with respect to
the ordered phases, the interplay of order parameters and surface segregation 
creates effective ordering surface fields\cite{ich1,ich3,dosch2,diehl1},
which may affect the critical behavior at the surface\cite{diehl1,upton}. 
In the case of a one component order parameter, such a field drives the
system from critical wetting to partial wetting.
When several order parameters are involved, this is not necessarily the
case\cite{helbing,gerhard1,gerhard2}. More subtle effects can lead to surface
order even at fully symmetric surfaces\cite{dosch4,schweika1,schweika3}.

Experimentally, surface induced disorder has been investigated at the (100) 
surface of Cu${}_3$Au\cite{sundaram,rae,alvarado,dosch3}. A number of studies 
have provided evidence that the order parameter right at the surface vanishes
continuously as the bulk transition is approached\cite{sundaram,rae,alvarado}, 
and established the relation with the existence of a disordered surface 
layer of growing thickness\cite{dosch3}.
The related case of ``interface induced disorder'' has been studied among
other in Cu-Pd, where the width of anti-phase boundaries was shown to diverge
logarithmically as the temperature of the transition to the disordered
phase was approached from below\cite{ricolleau}.

The first simulation studies of surface induced disorder in different 
systems have reproduced the continuous decrease of the surface order 
parameter at the bulk first order transition\cite{sundaram2,gerhard2,helbing}, 
and the logarithmic growth of a disordered layer near the surface\cite{helbing}.
A detailed study of surface induced disorder at the (111) surface of
CuAu has been published recently by Schweika {\em et al}\cite{schweika}. 
The critical behavior of various quantities has been analyzed, and critical 
exponents were found which agree well with the theory of critical wetting.  
Most notably, Schweika {\em et al} observe nonuniversal exponents, as
predicted by renormalization group theories of wetting phenomena\cite{bhl,fh}.
In contrast, Monte Carlo simulations of critical wetting in a simple Ising 
model have given results which were more consistent with mean field 
exponents\cite{KB1}. This latter finding has intrigued theorists for some 
time, and a number of theories have been put forward to account for the 
unexpected lack of fluctuation effects\cite{fjin,boulter,swain1}. 
The nonuniversality of the exponents observed by Schweika {\em et al} seems 
to indicate that the fluctuations are restored in the case of SID. 
Alternatively, it may also stem from a competition of different length 
scales associated with the local order parameter and the local 
composition\cite{gerhard1,gerhard2,hauge}.

In the present work, we study surfaces of a bcc-based alloy close to the
first order transition from the ordered DO${}_3$ phase to the disordered
phase. Our work is thus closely related to that of Schweika {\em et al}. 
It differs in that the order parameter structure in the bcc case
is much more complex than in the fcc-alloy:
Whereas only one (three dimensional) order parameter drives the transition
considered by Schweika {\em et al}, we have to deal with two qualitatively 
different order parameters, which are entangled with each other in a rather 
intriguing way.  In fact, we shall see that one of them behaves as expected 
from the theory of critical wetting, whereas the other exhibits different 
critical exponents, which do not fit into the current picture. 

A similar system has been investigated some time ago by 
Helbing {\em et al}\cite{helbing}. The systems studied there were 
rather small, and a detailed analysis of the critical behavior was 
not possible. Helbing {\em et al} report evidence for the presence of a 
logarithmically growing disordered layer at the (100)-surface as phase 
coexistence was approached. In retrospect, this result seems surprising,
since the (100) surface breaks the symmetry with respect to one of 
the order parameters, and we know nowadays that this nucleates an
ordering surface field. In order to elucidate the influence of this
ordering field in more detail, we have thus considered both
the (110) surface, which has the full symmetry of the bulk lattice,
and the (100) surface.

Our paper is organized as follows: In the next section, we provide some
theoretical background on the theory of wetting in systems with several
order parameters. Section \ref{model} is devoted to some general remarks on 
order-disorder transitions in bcc-alloys, and to the presentation of the model 
and the simulation method. Our results are discussed in section \ref{results}. 
We summarize and conclude in section \ref{summary}.

\section{Effective interface theory of surface induced disorder}
\label{wetting}

\subsection{General considerations}
\label{general}

We have already sketched one of the popular mean field approaches to wetting 
problems in the introduction. Since the bulk of the system is not critical, 
one can expect fluctuations to be negligeable for the most part. Only the
fluctuations of the local position $l(\vec{r})$ of the interface between the 
growing surface layer and the bulk phase remain important\cite{lkz}. 
As the interface moves into the bulk, capillary wave excursions of larger 
and larger wavelengths become possible. These introduce long-range 
correlations parallel to the surface, characterized by a correlation length 
$\xi_{\parallel}$ which diverges at wetting. 

In light of these considerations, fluctuation analyses often replace the 
Landau free energy functional (\ref{landau}) by an effective interface 
Hamiltonian\cite{bhl,lkz,fisher}
\begin{equation}
\label{hinter}
{\cal H} \{ l \} / k_B T 
= \int \!\! d \vec{r} \; \Big\{ \frac{1}{8 \pi \omega} 
(\nabla {l})^2 + V_0(l) \Big\}.
\end{equation}
Here all lengths are given in units of the bulk correlation length $\xi_b$
in the phase adsorbed at the surface,
the parameter $\omega$ is the dimensionless inverse of the 
interfacial tension $\sigma$
\begin{equation}
\omega = k_B T/4 \pi \sigma \xi_b{}^2,
\end{equation}
and the potential $V_0(l)$ describes effective interactions between the
interface and the surface. The wetting transition is thus identified with a 
depinning transition of the interface from the surface. 

In the linearized theory, the partition function of the Hamiltonian
(\ref{hinter}) is approximated by
\begin{equation}
{\cal Z} \approx
\int \! {\cal D} \{l\} \; e^{-\int \!\! d\vec{r}
\; (\nabla l)^2/8 \pi \omega } \;\;
\big[ \, 1 + \int \!\! d \vec{r} \; V_0(l)\, \big].
\end{equation}
It is convenient to switch from the real space $\vec{r}$ to the Fourier space
$\vec{q}$. The integration over short wavelength fluctuations with wavevector 
$|\vec{q}| > \lambda^{-1}$, where $\lambda$ is arbitrary, 
is then straightforward: One separates $l$ into a 
short wavelength part $\hat{l}(\vec{q}) = l(\vec{q})\;\theta (q-\lambda^{-1})$ 
and a long wavelength part $\overline{l} = l - \hat{l}$, and exploits
the relation $V_0(\overline{l}+x) = \exp[x \; d/dl] V_0(\overline{l})$, 
to obtain the unrescaled coarse grained potential \cite{lf}
\begin{eqnarray}
\overline{V}_{\lambda}(\overline{l}) & = &
\int \! {\cal D} \{\hat{l}\} \; \exp \Big[{-\frac{1}{4 \pi^2}\int \!\! d\vec{q}
\; \big\{ \frac{|\vec{q}\; \hat{l}|^2}{8 \pi \omega } 
+ \hat{l}\frac{d}{d l}} \big\} \Big] \;
V_0(\overline{l}) \nonumber\\
&=& \exp \Big[{\frac{\xi_{\perp,\lambda}^2}{2} \; (\frac{d}{dl})^2} \Big] 
V_{\lambda}(\overline{l}) 
\\
\mbox{with} && \xi_{\perp,\lambda}^2
=\frac{\omega}{\pi}\int_{q >1/\lambda}^{1/\Lambda} \!\!\!\ d \vec{q} \;
\frac{1}{q^2}
= 2 \omega \ln(\lambda/\Lambda),
\label{xip1}
\end{eqnarray}
where $\Lambda$ is a microscopic cutoff length. After rescaling 
$\vec{r}\to\vec{r}/\lambda$, $\overline{V}_{\lambda}(l)\to V_{\lambda}(l) 
=\lambda^{d-1}\overline{V}_{\lambda}(\overline{l}\lambda^{\zeta})$, and 
noting that the roughness exponent $\zeta$ is zero for capillary waves 
in $d=3$ dimensions, this can be rewritten as
\begin{equation}
\label{gauss}
V_{\lambda}(l) = 
\frac{\lambda^{2}}{\sqrt{2 \pi \xi_{\perp,\lambda}^2}} \int \!\! dh \; 
e^{-h^2/2 \xi_{\perp,\lambda}^2} \; V_0(l+h). 
\end{equation}
Renormalizing the potential $V_0(l)$ thus amounts to convoluting it
with a Gaussian of width $\xi_{\perp,\lambda}^2$\cite{bhl}, which is
the width of a free interface on the length scale $\lambda$ parallel to the 
interface. In the case of a bound interface, a natural choice 
for $\lambda$ is $\xi_{\parallel}$, the parallel correlation length of the 
interface. Since the remaining fluctuations after the renormalization should
be small on this length scale, the procedure can be made self consistent
by equating $\xi_{\parallel}$ with its mean field value 
\begin{equation}
\label{ddv}
4 \pi \omega \frac{d^2}{d l^2} V_{\lambda}(l)\Big|_{l=\langle l \rangle} =
(\xi_{\parallel}/\lambda)^{-2} = 1  \quad \mbox{at}
\quad \lambda = \xi_{\parallel}
\end{equation}
where the average position of the interface $\langle l \rangle$ is the 
position of the minimum of $V_{\lambda}(l)$. Note that the renormalized
free energy density per area $\xi_{\parallel}{}^2$ is of order unity. 
The singular part $F_s$ of the total interface free energy thus scales like 
$ F_s \propto \xi_{\parallel}^{-2}$.
From the renormalized Hamiltonian (\ref{hinter}),
\begin{equation}
{\cal H}_{\xi_{\parallel}} \{ l \} / k_B T = \frac{1}{4 \pi^2}
\int_0^{\xi_{\parallel}/\Lambda} \!\! d \vec{q} \; 
\frac{1}{8 \pi \omega} (q^2 + 1)\; \big|\, l(\vec{q})\, \big|^2
\end{equation}
we can now calculate the distribution probability to find the
interface at a position $h$,
\begin{equation}
\label{p1}
P(h) = \big\langle \delta[h - l(0)] 
\big\rangle_{{\cal H}_{{\xi}_{\parallel}^2}}
= \frac{1}{\sqrt{2 \pi \xi_{\perp}}} \; e^{-h^2/2 \xi_{\perp}^2}, 
\end{equation}
and the joint probability distribution that the interface is found at 
$h$ and $h'$ at two points separated by $\vec{r}$ from each other.
\begin{eqnarray}
\label{p2}
P^{(2)}(h,&&h',\vec{r}) = 
 \big\langle \delta[h - l(0)]\; \delta[h' - l(\vec{r})]
\big\rangle_{{\cal H}_{{\xi}_{\parallel}}} \nonumber\\
=&& \frac{1}{2 \pi \sqrt{g(0)^2-g(r)^2}} \nonumber\\
&& \times \;
\exp\big[- \frac{(h-h')^2}{4(g(0)-g(r))}
-\frac{(h+h')^2}{4(g(0)+g(r))}\big],
\end{eqnarray}
where 
\begin{equation}
\label{gr}
g(r) = \big\langle l(0) \; l(\vec{r}) 
\big\rangle_{{\cal H}_{{\xi}_{\parallel}}} 
 = \frac{\omega}{\pi} \int_0^{\xi_{\parallel}/\Lambda} \!\! 
\frac{d \vec{q}}{q^2+1} \; e^{i \vec{q}\vec{r}/\xi_{\parallel}}
\end{equation}
is the height-height correlation function of the interface and
\begin{equation}
\label{g0}
\xi_{\perp}^2 = g(0) \approx 2 \omega \ln (\xi_{\parallel}/\Lambda).
\end{equation}
An analogous expression has been derived by Bedeaux and Weeks for a free 
liquid-gas interface in a gravitational field\cite{bedeaux}. In three 
dimensions, the height-height correlation function for $r \gg \Lambda$ 
and $\xi_{\parallel} \gg \Lambda$ is a Bessel function $K$,
\begin{equation}
g(r) = 2 \omega \; K_0(r/\xi_{\parallel}).
\end{equation}

We assume that the average order parameter profile 
$\langle m(z) \rangle $ is given by the average over mean field order 
parameter profiles $m_{\mbox{\small bare}}(z-l)$, 
centered around the local interface positions 
$l$, which are distributed according to the distribution function $P(l)$. 
\begin{equation}
\label{maver}
\langle m(z) \rangle = \int \! dl \; P(l) \; m_{\mbox{\small bare}}(z-l)
\end{equation}
The distribution functions $P(h)$ and $P^{(2)}(h,h',r)$ can then be used
to calculate various characteristics of the profiles: 

For example, the effective width of the order parameter profile,
$W = 1/\big( 2 \; {\partial \langle m \rangle }/
{\partial z} \big)_{\langle l \rangle}$,
is broadened by $P(h)$ and diverges according to \cite{ich2,andreas}
\begin{equation}
\label{ew2}
W^2 \approx W_0{}^2 + \frac{\pi}{2} \; \xi_{\perp}{}^2
\end{equation}
where $W_0$ denotes the ``intrinsic width'' of the mean field profile,
$W_0 = 1/(2 \; dm_{\mbox{\small bare}}/dz)|_{z=0}$. 

Another quantity of interest is the layer-layer susceptibility, which
describes the order parameter fluctuations at a given distance from
the surface,
\begin{equation}
\label{chinn0}
\chi_{zz} = \int \! d \vec{r} \; \Big\{\big\langle m(0)m(\vec{r})\big\rangle_z
- \big\langle m \big\rangle^2_z
\Big\}.
\end{equation}
Since it has the dimension of a square length, one deduces immediately that
$\chi_{zz}$ scales like $\xi_{\parallel}{}^2$ in the interfacial region.
For a more detailed analysis, we rewrite $\chi_{zz}$ as 
\begin{eqnarray}
\chi_{zz} &=& 
\int \! d \vec{r} \; 
\int \! dh \; dh' \; m_{\mbox{\small bare}}(z-h) 
\; m_{\mbox{\small bare}}(z-h') \nonumber \\
& &\times
\int \! d \vec{r} \; 
\big\{{P^{(2)}(h,h',r)} - {P(h)\; P(h')} \big\},
\end{eqnarray}
expand the joint probability $P^{(2)}(h,h',r)$ in powers of
$\Delta(r) = g(r)/\xi_{\perp}^2 
= K_0(r/\xi_{\parallel})/\ln(\xi_{\parallel}/\Lambda)$, 
\begin{eqnarray}
P^{(2)}&&(h,h',r)  =  P(h)\;  P(h')\; \Big\{ \;
 1 + \frac{h \; h'}{\xi_{\perp}^2} \Delta(r) \nonumber\\
 && + \; \frac{1}{2} \: \big[ 1 - \frac{h^2 + h'^2}{\xi_{\perp}^2} + 
\frac{h^2 h'^2}{\xi_{\perp}^4} \big] \Delta(r)^2 + \cdots \Big\}, 
\end{eqnarray}
and recall $\int \! dr\,r\,K_0(r)=1$ and $\int \! dr\,r\,K_0(r)^2 = 1/2$.
If the intrinsic width of the profile $m_{\mbox{\small bare}}(z)$ is small 
compared to $\xi_{\perp}$, the intrinsic profile can be approximated by a simple
step profile in the interfacial region, 
$m_{\mbox{\small bare}}(z) = m_b \theta(z)$, 
where $m_b$ is the bulk order parameter. One then obtains
\begin{eqnarray}
\chi_{zz} &=& m_b{}^2 \xi_{\parallel}{}^2
e^{- (z-\langle l \rangle) ^2/\xi_{\perp}^2} \nonumber\\
\label{chizz}
&&\times \; \Big ( \, \frac{2 \omega}{\xi_{\perp}{}^2} 
+ \big(\frac{2 \omega}{\xi_{\perp}{}^2}\big)^2 \;
\frac{(z-\langle l \rangle)^2}{4 \xi_{\perp}{}^4} 
+ \cdots \Big).
\end{eqnarray}

So far, these results are valid for infinite systems.
The restriction to finite lateral dimension $L$ affects the interface 
distribution $P(h)$ (\ref{p1}) in two ways: It introduces a lower cutoff 
$\xi_{\parallel}/L$ in the integrals over $\vec{q}$ ({\em e.g.}, (\ref{gr})), 
and the mean position of the interface (the zeroth mode) is no longer fixed at 
the minimum of the renormalized potential, but distributed according to
$\exp[-L^2 V_{\xi_{\parallel}}(h)]$. The width of the distribution
function $P(h)$ is now given by
\begin{eqnarray}
\label{xipfss}
\xi_{\perp}^2 &=& \frac{\omega}{\pi} 
\int_{\xi_{\parallel}/L}^{\xi_{\parallel}/\Lambda} \!\! \frac{d \vec{q}}{q^2+1} 
+ L^2 \; \frac{d^2 V_{\xi_{\parallel}}(h)}{d h^2} \Big|_{h = \langle l \rangle}
\nonumber\\
&=& 2 \omega \ln(\frac{\xi_{\parallel}}{\Lambda}) 
- \omega \ln(1+(\frac{\xi_{\parallel}}{L})^2) + 
4 \pi \omega \; (\frac{\xi_{\parallel}}{L})^2.
\end{eqnarray}

\subsection{Bare and renormalized effective interface potential}
\label{special}

We shall now apply these general considerations to a specific potential 
$V_0(l)$, designed to describe systems with short range interactions and 
several order parameters and nonordering densities. Effective interface 
potentials for systems with two order parameters have been derived by 
Hauge\cite{hauge} and Kroll and Gompper\cite{gerhard1}. Their approach can 
readily be generalized to the case of arbitrary many order parameters and 
nonordering densities. We choose the coordinate system in the order parameter 
and density space $\{ \bf m \}$ such that ${\bf m} = 0$  in the phase which 
wets the surface, and that the coordinate axes $m_i$ point in the 
directions of the principal curvatures of the bulk free energy function
$f_b({\bf m})$. Close to this phase, $f_b$ can then be approximated by the
quadratic form
\begin{equation}
\label{fb}
f_b({\bf m}) =  \frac{g}{2} \sum_i \frac{1}{\lambda_i{}^2} m_i{}^2 + \mu,
\end{equation}
where $\mu$ is the field which drives the system from coexistence, and the
$\lambda_i$ have the dimension of a length. We number the coordinate axes $i$ 
($i \ge 0$) such that the $\lambda_i$ are arranged in descending order. The 
largest of these dominates the correlations at large distances and is thus
the correlation length $\xi_b$, {\em i.e.}, $\lambda_0 = \xi_b \equiv 1$ in 
our units. The surface contribution has the form
\begin{equation}
\label{fs}
f_s({\bf m}) = \sum_i h_{i,1} m_i + \frac{1}{2}\sum_{ij} c_{ij} m_i m_j.
\end{equation}
Following Hauge and Kroll/Gompper, we now assume that the actual profile 
from the adsorbed phase to the bulk phase is close to the profile of a free 
interface between these two phases. Close to the surface region, we 
thus approximate the former by the test function 
\begin{equation}
m_i(z) = v_i \exp{(z-l)/\lambda_i}
\end{equation}
(at $z \ll l$), where $l$ denotes the position of the effective interface. 
Inserting this into eqn. (\ref{landau}) with (\ref{fb}) and (\ref{fs}),
we obtain the effective interface potential
\begin{equation}
\label{vbare}
V_0(l) = \sum_i a_i \: e^{-l/\lambda_i} 
+ \sum_{ij} b_{ij} \: e^{-l (1/\lambda_i + 1/\lambda_j)} + \mu l
\end{equation}
for $l \gg  0$, with $a_i = h_{i,1} v_i$ and 
$b_{ij} = \frac{1}{2}(c_{ij} - g \delta_{ij}/\lambda_i) v_i v_j$. 
This expression is of course only valid for large $l$. Notably, it fails
at $l=0$, since the true potential $V_0(l)$ must diverge there.
We shall suppose that the leading term $b_{00} \equiv b$ in the second sum is 
positive and dominates over the more rapidly decaying terms, and disregard the
latter in the following.

At $\omega = 0$ (or in mean field approximation), the interface ist flat, 
and its position is given by the minimum of $V_0(l)$. At nonzero $\omega$,
the potential has to be renormalized as described in the previous section.
Now the renormalization is straightforward if the fluctuations are 
sufficiently small that the interface position $\langle l \rangle$ at
wetting is well in the asymptotic tail of the potential (weak fluctuation
limit). According to a criterion introduced by 
Br\'ezin, Halperin, and Leibler\cite{bhl}, this is true as long as 
$
\int_0^{\infty} dl\:  e^{-(l-\langle l \rangle)^2/2 \xi_{\perp}^2} V_0(l)
\approx
\int_{-\infty}^{\infty} dl \: 
e^{-(l-\langle l \rangle)^2/2 \xi_{\perp}^2} V_0(l),
$ {\em i.e.}, 
\begin{equation}
\label{wfl}
2 \xi_{\perp}^2 - \langle l \rangle < 0  
\qquad  \mbox{and} \qquad
\xi_{\perp}^2/\lambda_i - \langle l \rangle < 0
\end{equation}
for all $\lambda_i$. For $\lambda_i > 1/2$, the first inequality enforces
the second one. In a system with one order parameter, it leads to the 
well-known inequality $\omega < 1/2$\cite{bhl,fh}. As we shall see shortly, this
condition is also sufficient to ensure the validity of the weak fluctuation 
limit in a system with several order parameters. Since $\omega$ in our 
simulations turns out to be much smaller than 1/2, we shall not discuss the 
other regimes in the present paper.

In the weak fluctuation limit, the renormalized potential takes the form
\begin{equation}
\label{vxi}
\frac{V_{\xi_{\parallel}}(l)}{\xi_{\parallel}^2}
= \sum_{i: \lambda_i < 1/2} a_i \; e^{-l/\lambda_i} \; 
(\frac{\xi_{\parallel}}{\Lambda})^{\omega/\lambda_i^2} + 
b \; (\frac{\xi_{\parallel}}{\Lambda})^{4 \omega} e^{-2 l} + 
\mu l.
\end{equation}
The cutoff parameter $\Lambda$ is of the order of the correlation length,
$\Lambda \approx \xi_b = 1$, and will be dropped hereafter.

\subsection{Free energy scaling}
\label{scaling}

Now our task is to determine $\xi_{\parallel}$ self consistently by use 
of eqn. (\ref{ddv}), which will yield the scaling behavior of the singular
part of the surface free energy, $F_s \propto \xi_{\parallel}^{-2}$.
Before generalizing to several order parameters, we shall briefly discuss
the situation in a system with only one length scale $\lambda_0$. 
The formal alikeness of the more general theory with this often discussed 
special case can thus be highlighted. Moreover, many of the results 
derived for one order parameter carry over directly to the case of several 
order parameters.

In a system with one order parameter, the singular free energy has the scaling 
form
\begin{equation}
\label{fs0}
F_s \propto \xi_{\parallel}{}^{-2} = {8 \pi \omega} \; {\mu} \; 
f(\Phi_0),
\end{equation}
where the scaling function $f(\Phi_0)$ depends on the dimensionless parameter
\begin{equation}
\Phi_0 = C_0 \; \mu^{(\omega-1)/2} \; a_0 
\qquad \mbox{with} \qquad
C_0 = \sqrt{(8\pi \omega)^{\omega}/2 b}.
\end{equation}
Depending on the value of $\Phi_0$, one can distinguish between different
regimes:
\begin{eqnarray}
\lefteqn{\Phi_0 \gg 1:} \hspace{1.7cm}
&& f(\Phi_0) = 1/2 \; g_1 \! \big( 2^{\omega}\Phi_0^{-2} \big) 
\nonumber \\
   \mbox{with} \quad && 
g_1(x) \approx 1 + x - (2+\omega) x^2 + \cdots  \nonumber\\
 \lefteqn{\mbox{(complete wetting)}}  \\
\lefteqn{|\Phi_0| \ll 1:} \hspace{1.7cm}
&& f(\Phi_0) \approx 1 - \frac{1}{2}\Phi_0
+ \frac{2+\omega}{8} \Phi_0^2 + \cdots  \nonumber \\
 \lefteqn{\mbox{(critical wetting, field like)}} \\
\lefteqn{\Phi_0 \ll -1:} \hspace{1.7cm}  
&& f(\Phi_0) = (\Phi_0^2/2)^{1/(1-\omega)} \; 
g_2 \big((\Phi_0^2/2)^{-1/(1-\omega)}\big)  \nonumber \\
\mbox{with} \quad && 
g_2(x) \approx  \; 1 + \frac{3 x}{2(1-\omega)}  - 
\frac{(2 + 7 \omega) x^2}{8(1-\omega)^2}  + \cdots  \nonumber\\
 \lefteqn{\mbox{(partial wetting)}} 
\end{eqnarray}
The point $\Phi_0=0$ is the critical wetting point. If one approaches this
point from the partial wetting side $a_0 \to 0^-$ on the coexistence line 
$\mu = 0$, the parallel correlation length $\xi_{\parallel}$ diverges with
the well-known nonuniversal exponent
\begin{equation}
\xi_{\parallel} = (2 \pi \omega/b)^{-1/2(1-\omega)} \; (-a_0)^{-1/(1-\omega)},
\end{equation}
and the distance between the average position of the interface and the surface
diverges asymptotically like
\begin{equation}
\label{ell0}
\langle l \rangle \to - (1 + 2 \omega)/(1+\omega)  \; \ln (-a_0).
\end{equation}

The relevant regime for most cases of surface induced disorder is however the 
critical wetting regime, where the critical wetting point is approached under
a finite angle to the coexistence line in $(a_0, \mu)$ space.
Here the parallel correlation length $\xi_{\parallel}$ scales like
\begin{equation}
\label{xipar}
\xi_{\parallel} = \frac{1}{\sqrt{8 \pi \omega}} \; \mu^{-\nu_{\parallel}} 
\qquad \mbox{with} \qquad \nu_{\parallel} = 1/2. 
\end{equation}
as $\mu$ approaches zero, the width of the interface diverges with
\begin{equation}
\label{xiperp}
W^2 \to \frac{\pi}{2} \xi_{\perp}^2 = - \frac{\pi}{2} \omega \ln (\mu),
\end{equation}
and its average position with
\begin{equation}
\label{ell1}
\langle l \rangle \approx - (\omega + 1/2) \; \ln (\mu).
\end{equation}

These results can be used to derive the
layer-bulk susceptibility of the order parameter in the interfacial region
\begin{equation}
\label{chil}
\chi_{0,\langle l \rangle} 
= \frac{\partial \langle m_0 \rangle }{\partial \mu} \Big|_{\langle l \rangle}
\propto -\frac{\partial \langle m_0 \rangle}{\partial z} 
\Big|_{\langle l \rangle}\;
  \frac{\partial \langle l \rangle} {\partial \mu}
\propto \frac{1}{\mu \sqrt{ln (\mu)}}.
\end{equation}
In the step approximation $m_{0,\mbox{\small bare}}(z) = m_{0,b} \theta(z)$,
the layer-bulk susceptibility in the interfacial region can be calculated in 
more detail:
\begin{equation}
\label{chiz}
\chi_{0,z} = \frac{m_{0,b}}{\sqrt{2 \pi} \, \xi_{\perp} \, \mu} \;
e^{-(z-\langle l \rangle)^2/2 \xi_{\perp}{}^2}
\big( \omega +\frac{1}{2} - \frac{z-\langle l \rangle}{2 \ln \mu}\big).
\end{equation}
It has a slightly asymmetric peak of width $\xi_{\perp}$ at 
$z = \langle l \rangle$, the height of which scales like $1/\mu$. 

The layer-layer susceptibility could already be derived in the previous
section. It also has a peak at the interface, which is however a factor of
$\sqrt{2}$ narrower. Its height scales like
\begin{equation}
\label{chill}
\chi_{\langle l \rangle\langle l \rangle}  
\propto \xi_{\parallel}{}^2/\xi_{\perp}{}^2
\propto -1/(\mu \; \ln(\mu)).
\end{equation}

Next we determine the critical behavior of the order parameter at the 
surface, $m_{0,1}$, 
\begin{equation}
\label{m1_1}
m_{0,1} \propto - \frac{\partial F_s}{\partial h_{0,1}}
\propto - \frac{\partial\xi_{\parallel}^{-2}}{\partial a_0}
\propto \mu^{\beta_{0,1}},
\qquad \beta_{0,1} = \frac{1+\omega}{2} 
\end{equation}
It will prove useful to rederive the exponent $\beta_{0,1}$ 
in an alternative way: The surface order parameter in mean
field theory is given by $m_{\mbox{\small bare}}(0)=m_b\exp(-l/\lambda_0)$ 
Averaging the profile according to eqn. (\ref{maver}) yields
\begin{equation}
\label{m1_2}
m_{0,1}  =  m_b \langle e^{-l/\lambda_0} \rangle_{P(l)} 
= m_b e^{- \langle l \rangle/\lambda_0 + \xi_{\perp}^2/2 \lambda_0^2}.
\end{equation}
After inserting $\lambda_0 = 1$ and using eqns. (\ref{ell1}) and 
(\ref{xiperp}), one recovers the power law of eqn. (\ref{m1_1}) 
with the same exponent $\beta_{0,1}$.
The approach has the advantage that it allows for a straightforward 
calculation of finite size effects on surface critical behavior:
We simply replace the expression (\ref{xiperp}) for $\xi_{\perp}$ in the 
infinite system by eqn. (\ref{xipfss}) to obtain
\begin{equation}
\label{m1s1}
m_{0,1} \propto  m_b \; \mu^{\beta_{0,1}} \; 
\hat{M}_0( 8 \pi \omega \; \mu L^{1/\nu_{\parallel}} )
\end{equation}
with the scaling function
\begin{equation}
\label{m1s2}
\hat{M}_0(x) = \big( \frac{x}{x+1} \big)^{\omega/2} \;
e^{2 \pi \omega /x}.
\end{equation}

We are now ready to generalize these results to the case of several order 
parameters and nonordering densities. Formally, the theory turns out to remain 
very similar. The self consistent determination of $\xi_{\parallel}$ leads to a 
generalized version of the scaling form for the singular part of the surface 
free energy (\ref{fs0}),
\begin{equation}
\label{fs1}
F_s \propto \xi_{\parallel}{}^{-2} = {8 \pi \omega} \; {\mu} \; 
f(\{\Phi_i\}),
\end{equation}
where the scaling variables are
\begin{equation}
\Phi_i = C_i \; \mu^{(1-2 \lambda_i)(1-\omega/\lambda_i)/2 \lambda_i} \; a_i 
\end{equation}
\begin{displaymath}
\qquad \mbox{with} \qquad
C_i = (8\pi \omega)^{\omega/2 \lambda_i^2 \cdot (2 \lambda_i - 1)}
(2 b)^{-1/2 \lambda_i}.
\end{displaymath}

As in the one-order parameter case, we have to distinguish between different
regimes depending on the values of the scaling variables. 

\subsection{Symmetry preserving and symmetry breaking surfaces}
\label{surface}

Let us first assume that the effect of nonordering densities can be
disregarded ({\em e.g.}, because the associated length scales are small,
$\lambda_i < 1/2$), and consider the case of a symmetry preserving surface.
No ordering surface fields are then present, {\em i.e.}, 
$a_i \propto h_i = 0$ for all contributions $i$. The system is thus in a
``multicritical wetting regime'', where $|\Phi_i | \ll 1$ for all $i$, 
and the scaling function can be expanded as
\begin{equation}
f(\{ \Phi_i \}) = 1 - \sum_i \Phi_i \: \frac{2 \lambda_i - 1}{2 \lambda_i^2}
 + \cdots
\end{equation}
The effective interface position $\langle l \rangle$, and the correlation
length $\xi_{\parallel}$ are given by eqns. (\ref{ell1}) and (\ref{xipar})
as in the case of normal critical wetting. Hence all the results related to
interfacial properties, such as the interfacial width, the interfacial
layer susceptibilities etc., remain unchanged. In particular, the
criterion for the validity of the weak fluctuation limit is still
$\omega < 1/2$ (from eqns. (\ref{wfl}), (\ref{xiperp}) and (\ref{ell1})).
The surface order parameters obey the power law
\begin{equation}
m_{i,1} \propto - \frac{\partial \xi_{\parallel}^{-2}}{\partial a_i}
\propto \mu^{\beta_{i,1}}, \quad
\beta_{i,1} = \frac{1}{2 \lambda_i} 
+ \frac{\omega}{2 \lambda_i^2} (2 \lambda_i - 1).
\end{equation}
Following the lines of (\ref{m1_2}), one also obtains the finite size scaling 
function
\begin{equation}
\label{mscal}
\hat{M}_i(x) = \big( \frac{x}{x+1} \big)^{\omega/2 \lambda_i^2} \;
e^{2 \pi \omega /x \lambda_i^2}.
\end{equation}

A whole sequence of surface exponents is thus predicted, one for each order 
parameter. In practice, however, one hardly ever measures only one "pure" order 
parameter $m_i$. Instead, one expects to observe some combination of 
contributions with different exponents $\beta_{i,1}$, which will be dominated 
by the leading exponent $\beta_{0,1} = (\omega + 1)/2$ in the asymptotic 
limit $\mu \to 0$.

The situation changes when at least one of the $a_i$ becomes nonzero at
coexistence. This is the case, {\em e.g.}, at a symmetry breaking surface, 
where one or several surface fields become nonzero, or even at a symmetry
preserving surface if the length scale associated with a nonordering
density exceeds half the bulk correlation length, $\lambda_i > 1/2$.

Let $a_J e^{-l/\lambda_J}$ be the leading nonvanishing term in the potential 
(\ref{vbare}). As one approaches coexistence, $\mu \to 0$, the scaling variable 
$\Phi_J$ increases and one eventually enters a regime $|\Phi_J| \gg 1$. 
For negative $a_J$, ($\Phi_J \ll -1$), the wetting becomes partial, {\em i.e.}, 
no surface induced disordering takes place.
For positive $a_J$, ($\Phi_J \gg 1$), different scenarios are possible,
depending on the sign and the amplitude of the higher order terms $a_i$,
($i>J$) in eqn. (\ref{vxi}). If they are positive or sufficiently small, 
such that
\begin{equation}
\label{small}
|a_i a_J^{-\lambda_J/\lambda_i}| \ll 1,
\end{equation}
the disordered phase wets the surface. The effective interface 
position $\langle l \rangle$ diverges asymptotically like
\begin{equation}
\label{ell2}
\langle l \rangle \approx - \lambda_J (1 + \omega/2 \lambda_J^2) \; \ln (\mu),
\end{equation}
the parallel correlation length scales like
\begin{equation}
\xi_{\parallel} = \sqrt{\lambda_J/4 \pi \omega \mu},
\end{equation}
and the scaling function in eqn. (\ref{fs1}) takes the form
\begin{displaymath}
f(\{\Phi_i\}) =
\frac{1}{2 \lambda_J}
\Big( 1 + \sum_i \Phi_i \Phi_J^{-\lambda_J/\lambda_i} K_J(\lambda_i) \Big.
\end{displaymath}
\begin{equation}
\label{fs2}
\qquad \Big. + \; \frac{1}{2} \Phi_J^{-2 \lambda_J} K_J(\frac{1}{2}) \; \Big).
\end{equation}
with
\begin{displaymath}
 K_J(\lambda_i) = \frac{\lambda_J^{\lambda_J/\lambda_i}}{\lambda_i} \;
(\frac{\lambda_J}{\lambda_i} - 1) \;
(2 \lambda_J)^{\omega/2 \lambda_i^2 \cdot (1-\lambda_i/\lambda_J)}.
\end{displaymath}
According to eqn. (\ref{wfl}), the weak fluctuation regime here is bounded by 
$\omega < 2 \lambda_J^2$, thus encompassing the regime $\omega < 1/2$.

The criterion (\ref{small}) is motivated as follows: If one of the higher
order $a_i$ is negative and large, the interface potential 
$V_{\xi_{\parallel}}(l)$ may exhibit a second minimum closer to the
surface, which competes with the minimum at large $l$ and may prevent the
formation of an asymptotically diverging wetting layer. 
The inspection of the free energy scaling function (\ref{fs2}) reveals 
that the transition to such a partial wetting regime is appropriately described
in terms of the combined scaling variable
\begin{displaymath}
\tilde{\Phi}_{i,J} = \Phi_i \Phi_J^{-\lambda_J/\lambda_i}
\propto
a_i a_J^{-\lambda_J/\lambda_i} 
\mu^{(\lambda_J/\lambda_i - 1)(1- \omega/(2\lambda_i\lambda_J))}.
\end{displaymath}
This quantity has to be large at the point $\mu_0$ where the one minimum
of $V_{\xi_{\parallel}}(l)$ splits up in two.
The condition (\ref{small}) ensures that $\tilde{\Phi}_{i,J}$ is small
for all $\mu$.

The wetting is critical with respect to all order parameters $m_i$ with length 
scales $\lambda_i$ larger than  $\lambda_J$. As coexistence is approached,
they vanish at the surface according to the power law
\begin{equation}
\label{b1_2}
m_{i,1} \propto - \frac{\partial \xi_{\parallel}^{-2}}{\partial a_i}
\propto \mu^{\beta_{i,1}}, \quad
\beta_{i,1} = \frac{\lambda_J}{\lambda_i}
+ \frac{\omega}{2 \lambda_i^2} (\frac{\lambda_i}{\lambda_J} - 1).
\end{equation}
The finite size scaling function $\hat{M}_i(x)$ is again given by 
(\ref{mscal}), with the scaling variable $x = 4 \pi \omega \mu L^2 /\lambda_J$.
Note that the exponents $\beta_{i,1}$ are nonuniversal even in the mean field 
limit ($\omega = 0$). This remarkable effect has first been discovered by 
Hauge\cite{hauge} and later studied by Kroll/Gompper in an fcc Ising 
antiferromagnet using a mean field approximation\cite{gerhard1}, 
Monte Carlo simulations, and a linear renormalization group study similar to 
the one presented here\cite{gerhard2}. However, $\langle l \rangle$ in this
work is taken from eqn. (\ref{ell0}) rather than determined self consistently,
hence the resulting critical exponents differ somewhat from those calculated
here. As in the case of the symmetry preserving surface, a whole set of
exponents is predicted by eqn. (\ref{b1_2}). In the asymptotic limit
$\mu \to 0$, however, the surface behavior is expected to be governed
by the leading exponent
\begin{equation}
\label{b1_3}
\beta_{0,1} = \frac{\xi_b}{\lambda_J}
+ \frac{\omega}{2} (\frac{\xi_b}{\lambda_J} - 1).
\end{equation}
We have reinserted the bulk correlation length $\xi_b \equiv 1$ here.

Finally, we discuss the critical behavior of the surface susceptibilities.
The corresponding critical exponents can be shown to obey simple scaling laws. 
In the case of the surface-bulk susceptibility, the relation follows trivially:
\begin{equation}
\label{chi1}
\chi_{i,1} \propto \frac{\partial m_{i,1}}{\partial \mu} 
\propto \mu^{-\gamma_{i,1}}, 
\qquad \gamma_{i,1} = 1 - \beta_{i,1}.
\end{equation}
In the case of the surface-surface susceptibility, it depends on the regime
under consideration. In the ''critical wetting regimes'' discussed here,
the free energy scaling function $f$ can be expressed as a Taylor series
in powers of the scaling variables $\Phi_i$ or $\tilde{\Phi}_{i,J}$, 
respectively, and one
obtains
\begin{equation}
\label{chi11}
\chi_{i,11} \propto \frac{\partial m_{i,1}}{\partial h_{i,1}} 
\propto \mu \frac{\partial^2 f}{\partial a_i{}^2}
\propto \mu^{-\gamma_{i,11}}, 
\qquad \gamma_{i,11} = 1 - 2 \beta_{i,1}.
\end{equation}
The dominating exponents in the asymptotic limit are 
$\gamma_{0,1}$ and $\gamma_{0,11}$.

\section{Modeling order-disorder transition in bcc-alloys}
\label{model}

\begin{figure}[b]
\noindent
\hspace*{0.5cm}\fig{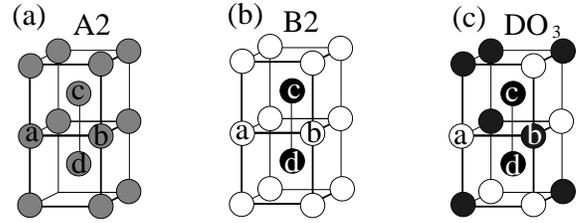}{75}{35}{
\vspace*{0.5cm}
\caption{bcc lattice with (a) disordered A2 structure (b) B2 structure, 
and (c) DO${}_3$ structure. Also shown is the assignment of 
sublattices a,b,c,d.}
\label{phases}
}
\end{figure}

Fig. \ref{phases} shows some typical structures of binary (AB) bcc-alloys 
({\em e.g.}, FeAl \cite{kubaschewski}). It is useful to divide the bcc lattice 
into four fcc-sublattices a--d as indicated in the Figure. 
The phase transitions are then conveniently 
described in terms of a set of order parameters\cite{duenweg}

\begin{eqnarray}
\psi_1 &=& \big( c_a + c_b - c_c - c_d) \nonumber \\
\psi_2 &=& \big( c_a - c_b + c_c - c_d)  \\
\psi_3 &=& \big( c_a - c_b - c_c + c_d), \nonumber
\end{eqnarray}
where $c_{\alpha}$ denotes the composition on the sublattice $\alpha$,
{\em i.e.}, the average concentration of one component $A$ there. 
In the disordered phase, all sublattice compositions are equal and
these order parameters vanish. The B2 phase is characterized by 
nonzero $\psi_1$, and the DO${}_3$ phase in addition by nonzero
$\psi_2 = \pm \psi_3$.  By symmetry, physical quantities have to be 
invariant under sublattice exchanges $(a \leftrightarrow b)$, 
$(c \leftrightarrow d)$, and $(a,b) \leftrightarrow (c,d)$. 
The leading terms in a Landau expansion of the free energy $F$ thus read
\begin{eqnarray}
\label{flandau}
 F &=& F_0 + A_1 \psi_1^2 + A_2 (\psi_2^2 + \psi_3^2) 
  + B \; \psi_1 \psi_2 \psi_3 \\
&&  + C_1 \psi_1^4 + C_2 (\psi_2^4 + \psi_3^4) 
  + C_3 \psi_2^2 \psi_3^2 + C_4 \psi_1^2 (\psi_2^2+\psi_3^2),\nonumber
\end{eqnarray}
We point out in particular the cubic term $B \psi_1 \psi_2 \psi_3$. 
It can be read in two ways. On the one hand, it describes how the B2-order 
influences the DO${}_3$ order: The order parameter $\psi_1 $ breaks the 
symmetry with respect to indi-
\noindent
vidual sign reversal of $\psi_2$ or $\psi_3$ 
and orients $(\psi_2,\psi_3)$ such that 
$\psi_2 = - \mbox{sign}(B \psi_1)\; \psi_3$. Conversely, one can interpret
the product $\psi_2 \psi_3$ as an effective ordering field acting on $\psi_1$. 
We shall come back to this point later.

At the presence of surfaces, the situation is even more complicated.
First, we can always expect that one component enriches at the surface, 
since there are no symmetry arguments to prevent that.
Even if no explicit surface field coupling to the total concentration $c$ is 
applied, the component which is in excess with respect to the ideal 
stoechiometry of the bulk phase ((3:1) in the DO${}_3$ phase) 
will segregate to the surface. Second, we have already mentioned 
that the Landau expansion of the surface free energy $f_s$ depends 
on the orientation of the surface\cite{ich1,diehl1}. 
The (110) surface has the same symmetry with respect to sublattice exchanges 
as the bulk, hence the Landau expansion of the surface free energy must have 
the form (\ref{flandau}). In case the order is sufficiently suppressed at the 
surface, one can thus hope to find classical surface induced disordering here.
In the case of the (100) surface, the symmetry with respect to the exchange 
$(a,b)\leftrightarrow (c,d)$ is broken. The surface enrichment
of one component then induces an effective ordering surface field, which 
couples to the order parameter 
$\psi_1$\cite{ich3}. 
Other ordering fields 
coupling to $\psi_2$ and $\psi_3$ are still forbidden by symmetry. The full 
spectrum of possible ordering surface fields is allowed in the case of
the (111) surface.

\begin{figure}
\noindent
\hspace*{1.5cm}\fig{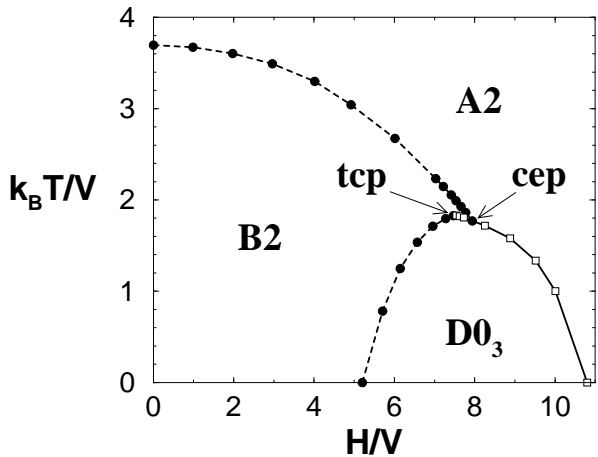}{65}{68}{
\vspace*{0.cm}
\caption{Phase diagram of our model in the $T-H$ plane. Solid lines mark first
order phase transitions, dashed lines second order phase transitions.
Arrows indicate the positions of a critical end point (cep) and a
tricritical point (tcp).
}
\label{phdiag}
}
\end{figure}

In order to model these phase transitions, we consider an Ising model
of spins $S_i = \pm 1$ on the bcc-lattice with antiferromagnetic interactions 
between up to next nearest neighbors,
\begin{equation}
{\cal H} =  V \sum_{\langle ij \rangle} S_i S_j
+ \alpha V \sum_{\langle \langle ij \rangle \rangle } S_i S_j 
- H \sum_i S_i.
\end{equation}
where the sum $\langle ij \rangle$ runs over nearest neighbour  
and $\langle \langle ij \rangle \rangle$ over next nearest neighbour pairs.
Spins $S=+1$ represent A-atoms and $S=-1$ B-atoms, hence the concentration
$c$ of A is related to the average spin $\langle S \rangle$ via
\begin{equation}
\label{cc}
c = (\langle S \rangle + 1)/2,
\end{equation}
and the field $H$ represents a chemical potential.
The parameter $\alpha = 0.457$ is chosen such that the highest temperature
which can still support a B2 phase is about twice as high as the highest 
temperature of the DO${}_3$ phase, like in the experimental phase diagram
of FeAl. The phase diagram of our model is shown in Figure \ref{phdiag}.

The surface simulations were performed in a 
$L \times L \times D$ geometry with periodic boundary conditions in the 
$L$ direction and free boundary conditions in the $D$ 
direction, varying $D$ from 100 to 200 and $L$ from 20 to 100. 
In order to handle systems of that size efficiently, we have 
developed\cite{frank} a multispin code\cite{bhanot}, which allowed to store 
the configurations bitwise instead of bytewise\cite{fn2}. 
Our Monte Carlo runs had total lengths of up to $2\cdot 10^6$ 
Monte Carlo sweeps.

\noindent 

\section{Simulation results}
\label{results}

We have studied (110) and (100) oriented surfaces at $T = 1 \: V/k_B$
close to the first order bulk transition between the ordered DO${}_3$ phase 
and the disordered A2 phase. The exact bulk transition point was
determined previously from bulk simulations by thermodynamic 
integration\cite{KB2}, $H_0/V = 10.00771(1)$\cite{frank}.
In the presence of such a high 
bulk field, the very top layer of a free (110) or (100) surface is completely 
filled with $A$ particles, {\em i.e.}, Ising spins $S=1$. Consequently, the 
order parameters $\psi_{\alpha}$ and the layer susceptibilities vanish there.
In the following, we shall generally disregard this top layer and 
analyze the profiles starting from the second layer. 

\subsection{(110) Surfaces -- DO${}_3$ order}

\begin{figure}
\noindent
\hspace*{0.5cm}\fig{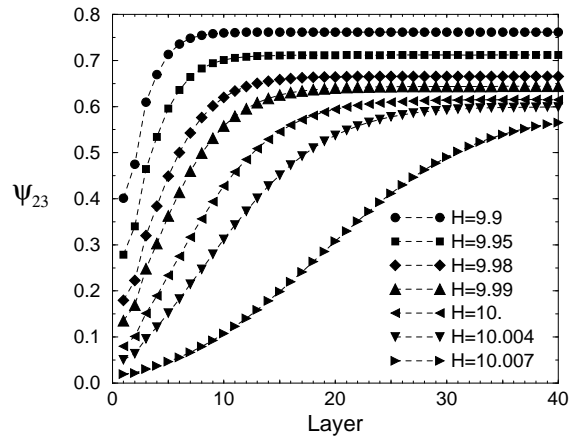}{75}{64}{
\vspace*{0.cm}
\caption{Profiles of $\psi_{23}$ near a (110) surface at temperature 
$T=1 \: V/k_B $ for different fields $H$ in units of $V$ as indicated. The bulk
transition is at $H_0/V = 10.00771(1)$.
Zeroth (top) layer is not shown ($\psi_{23}(0) \equiv 0$, see text).
}
\label{psi23}
}
\end{figure}

\begin{figure}
\noindent
\hspace*{1.cm}\fig{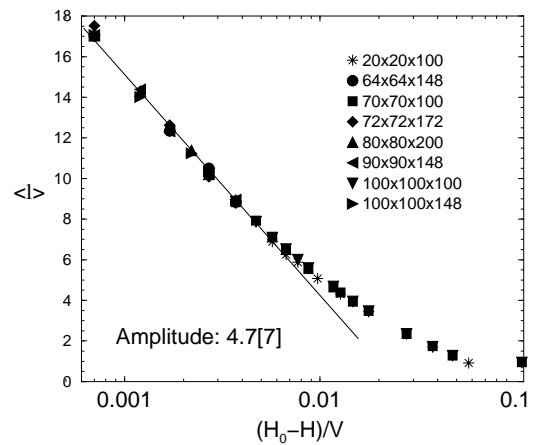}{70}{60}{
\vspace*{0.cm}
\caption{Position of the interface as estimated from the fit (65)
in units of (110) layers vs. $(H-H_0)/V$.}
\label{ll}
}
\end{figure}

\begin{figure}
\noindent
\hspace*{0.8cm}\fig{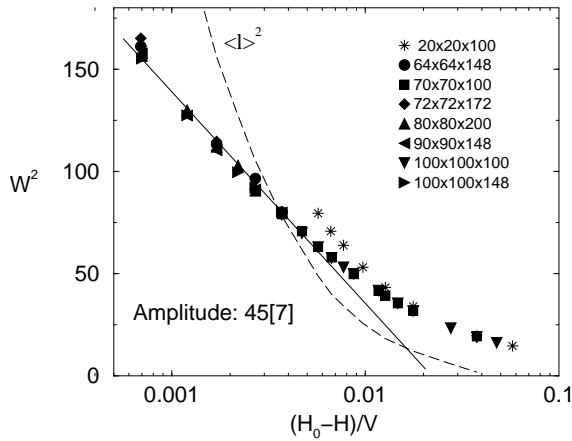}{74}{60}{
\vspace*{0.cm}
\caption{Squared interfacial width as estimated from the fit (65)
in units of (110) layers vs. $(H-H_0)/V$. Long dashed
line shows squared interface position $\langle l \rangle^2$ for
comparison.}
\label{w2}
}
\end{figure}

We begin with a detailed discussion of surface induced disordering
at (110) surfaces, {\em i.e.}, surfaces with the full symmetry of the bulk. 
Figure \ref{psi23} shows profiles of the order parameter of DO${}_3$ ordering
per site
\begin{equation}
\psi_{23} = \sqrt{(\psi_2{}^2 + \psi_3{}^2)/2}.
\end{equation}
One clearly sees how a disordered layer forms and grows in thickness as the 
bulk transition is approached. In order to extract an interface position 
$\langle l \rangle$ and an effective interfacial width $W$, 
we have fitted the profiles to a shifted tanh function
\begin{equation}
\label{fit}
\psi_{23}(n) = \psi_{23}^{\mbox{\tiny bulk}}
\Big( 1 + \exp \big[- 2 \; (z-\langle l \rangle)/W \: \big] \; \Big)^{-1}.
\end{equation}
The results are shown in Figs. \ref{ll} and \ref{w2}. Sufficiently close to 
the bulk transition, at $(H_0-H)/V < 0.005$, the data are consistent with the 
logarithmic divergence predicted by eqns. (\ref{ell1}) and (\ref{ew2}).
Intuitively, one would expect that an effective interface theory is only
applicable if $l > W$, 

\noindent
{\em i.e.}, the width of the interface is smaller than
the distance of the interface from the surface. Indeed, Fig. \ref{w2} shows
that the logarithmic behavior sets in approximately at the value of $H$ where 
$l$ begins to exceed $W$. The prefactors of the logarithms in Figs \ref{ll} 
and \ref{w2} are predicted to be $(r/2+\omega/r) \sqrt{2} \xi_b/a_0$ 
in the case of $\langle l \rangle$ (Fig. \ref{ll}), and 
$\pi \omega \xi_b^2/a_0^2$ in the case of $W^2$ (Fig. \ref{w2}), 
where $\xi_b$ is the bulk correlation length, $a_0$ the lattice constant, 
a factor $\sqrt{2}$ or $2$ accounts for the distance of (110) layers from 
each other in units of $a_0$, and the parameter 
$r = \mbox{max}(1,2 \lambda_J/\xi_b)$ depends on the length scale
$\lambda_J$ of composition fluctuations (see the discussion in section 
\ref{surface}). We shall see below that the surface data suggest 
$\beta_1 = r/2 + \omega (1/r - 1/2) = 0.618$. 
Inserting this result, one derives 
$ 4.5[7] < \xi_b/a_0 < 5.4[8]$ from Fig. \ref{ll}, and $\xi_b/a_0 > 7.8[8]$
from Fig. \ref{w2}. These values do not agree with each other within
in the statistical error; the interfacial width seems to decrease too 
fast as one moves away from coexistence. 
Yet the difference seems still acceptable, especially considering
how small the region of apparent logarithmic behavior is. It has been
observed in other systems\cite{andreas2}, that the vicinity of surfaces
also affects the intrinsic width $W_0$ of an interface.
Moreover, many non-diverging terms have been neglected in eqns. (\ref{ell1}) 
and (\ref{ew2}) which lead to systematic errors if one is not close enough 
to $H_0$. We note that $\xi_b$ seems rather large for a system which is not 
critical in the bulk. On the other hand, Fig. \ref{psi23} shows that the bulk 
order parameter $\psi_{23}$ decreases considerably as one approaches the 
phase transition point. This observations suggests that a critical point
is at least nearby, although preempted by the first order transition from 
the DO${}_3$ phase to the disordered phase. 

\begin{figure}
\noindent
\hspace*{0.5cm}\fig{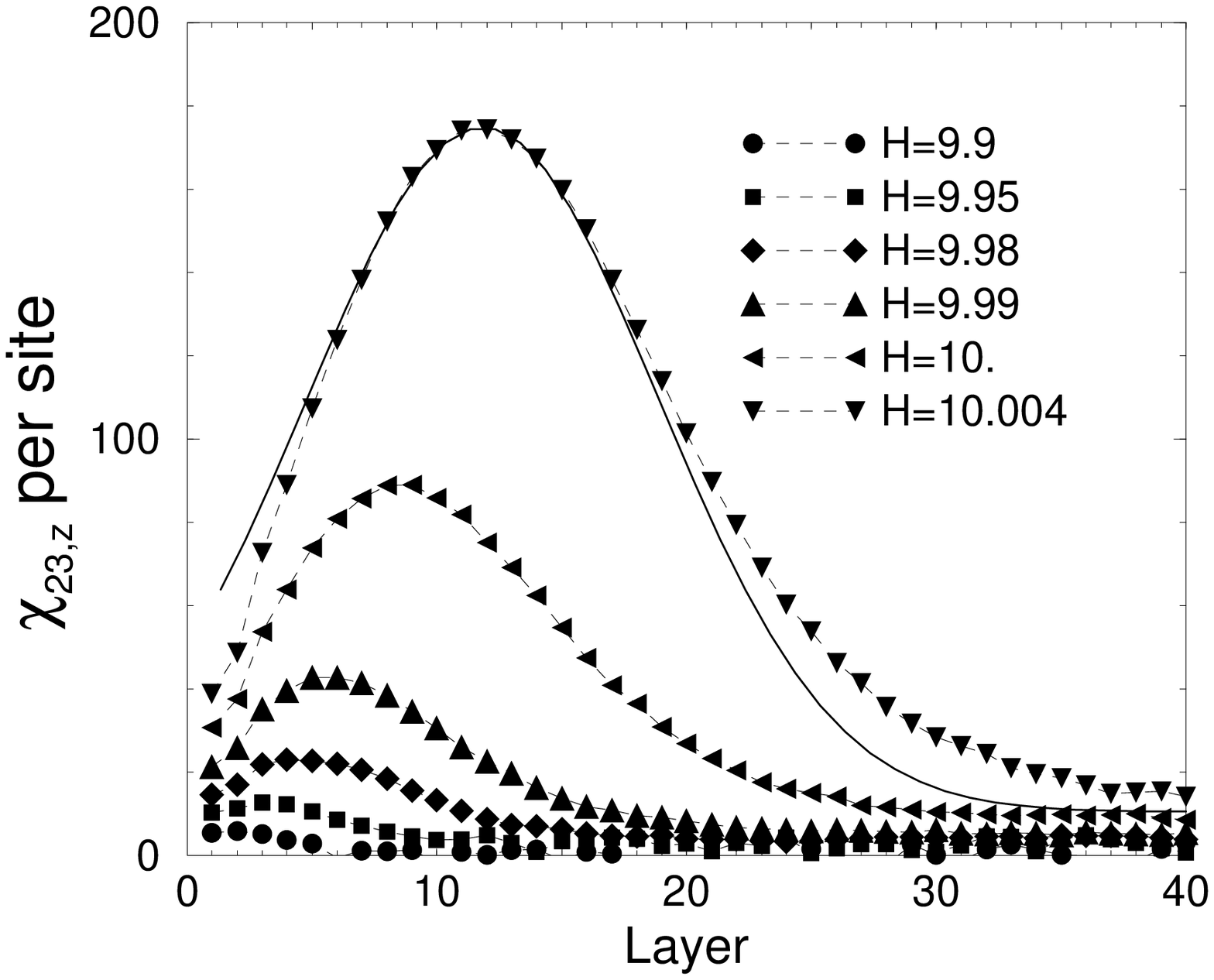}{70}{65}{ 
\vspace*{-60mm} 

(a) 

\vspace*{60mm}
}

\noindent
\hspace*{0.6cm}\fig{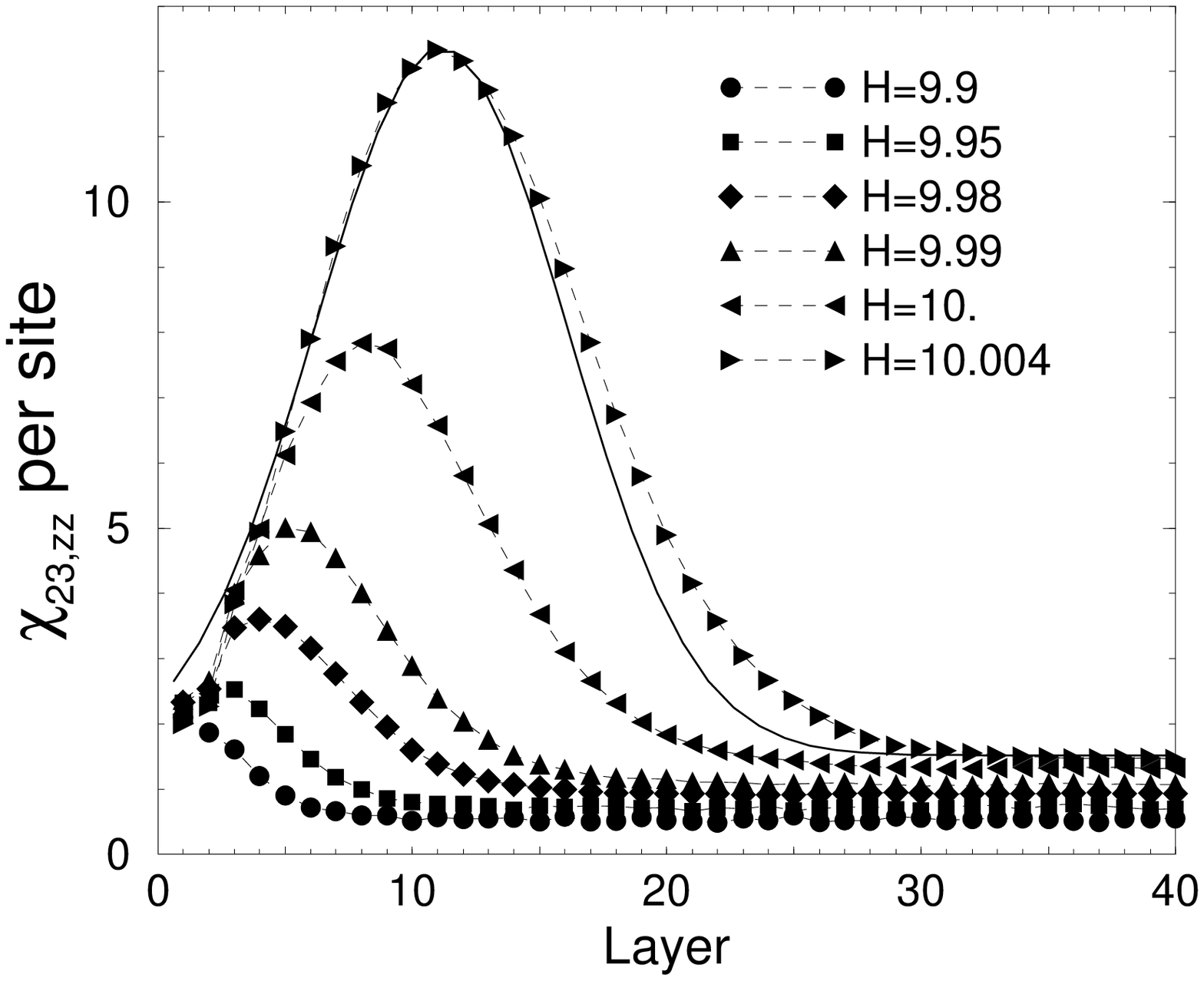}{68}{55}{
\vspace*{-60mm} 

(b) 

\vspace*{58mm}
\caption{
Profiles of the layer-bulk susceptibility $\chi_z$ (a) and
the layer-layer susceptibility $\chi_{zz}$ (b) per site of
the order parameter $\psi_{23}$ in units of $k_B T$, 
for different fields $H$ in units of $V$ as indicated.
Solid line shows the fit of a Gaussian of width 
(a) $\xi_{\perp}= (2/\pi)^{1/2} W$ and (b) $\xi_{\perp}/2^{1/2}$
to the profile corresponding to $H=10.004$. 
Zeroth (top) layer is not shown ($\chi(0) \equiv 0$, see text).
}
\label{chiprof}
}
\end{figure}

Next we consider the profiles of the layer susceptibilities of the
order parameter $\psi_{23}$. They can be determined from the simulation 
data by use of the fluctuation relations\cite{schweika}
\begin{eqnarray}
\chi_{z} &=& \frac{N_{\mbox{\tiny total}}}{k_B T}
 \Big( \langle \psi(z) \psi_{\mbox{\tiny total}} \rangle - 
\langle \psi(z) \rangle \: \langle \psi_{\mbox{\tiny total}} \rangle  \Big) \\
\chi_{zz} &=& \frac{N_{\mbox{\tiny layer}}}{k_B T}
 \Big( \langle \psi(z)^2 \rangle - \langle \psi(z) \rangle^2  \Big),
\end{eqnarray}
where $\psi$ is the order parameter under consideration, 
$N_{\mbox{\tiny layer}}$ denotes the number of sites in a layer, and
$N_{\mbox{\tiny total}}$ the total number of sites.
Fig. \ref{chiprof} shows that both the layer-bulk susceptibility $\chi_z$ and 
the layer-layer suszeptibility $\chi_{zz}$ exhibit the expected peak in
the vicinity of the interface (eqns. (\ref{chiz}) and (\ref{chizz})).
The centers of the peaks 
can be fitted nicely by Gaussians of width $\xi_{\perp}$ and 
$\xi_{\perp}/\sqrt{2}$, respectively, where 
$\xi_{\perp}$ is calculated from the width $W$ of the order parameter 
profile using $\xi_{\perp} = \sqrt{2/\pi} \, W$. The wings of the
peaks are not Gaussian any more, but asymmetric -- the layer susceptibilities
are enhanced at the bulk side of the interface, and suppressed at the
surface side. Such an asymmetry has been predicted qualitatively for $\chi_z$
in eqn. (\ref{chiz}), but not for $\chi_{zz}$ (cf. eqn. (\ref{chizz})). 
Even in the case of $\chi_z$, the observed asymmetry is so strong that it
cannot be brought into quantitative agree-

\noindent

\begin{figure}
\noindent
\hspace*{0.5cm}\fig{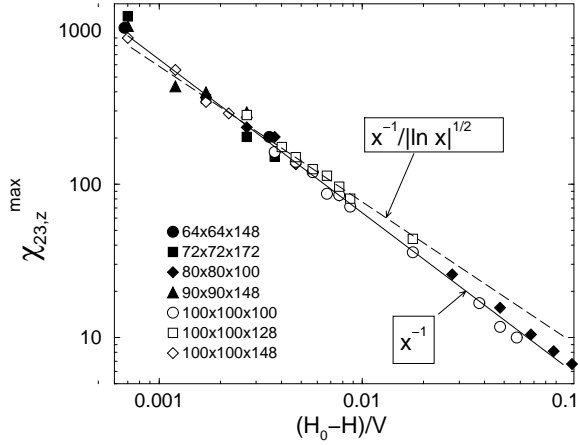}{75}{62}{ 
\vspace*{0.2cm}
\caption{
Maximum of the layer-bulk susceptibility $\chi_z$ per site of
the order parameter $\psi_{23}$ in units of $k_B T$ vs. $(H_0-H)/V$ 
for different system sizes as indicated.
Solid line shows a fit to a $(H_0-H)^{-1}$ behavior, 
and dashed line the same with logarithmic correction (see text).
}
\label{chin23m}
}
\end{figure}
\noindent

\begin{figure}
\noindent
\hspace*{0.5cm}\fig{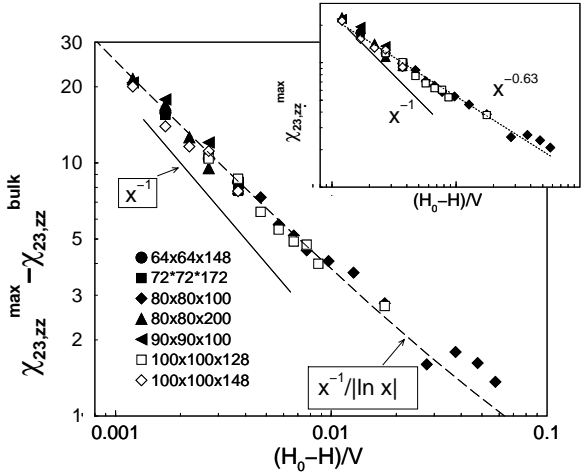}{75}{62}{
\vspace*{0.2cm}
\caption{
Maxima of the layer-layer susceptibility $\chi_{zz}$ per site of
the order parameter $\psi_{23}$ in units of $k_B T$ vs. $(H_0-H)/V$, 
for different system sizes as indicated.
Inset shows bare data, with a fit to a power law behavior with 
unknown exponent (dotted line). In the main plot, the bulk contribution 
to $\chi_{zz}$ has been subtracted. Solid line indicates the slope of 
$(H_0-H)^{-1}$, and dashed line the whole theoretical prediction including
the logarithmic correction.
}
\label{chinn23m}
}
\end{figure}
\noindent
ment with the theory. We recall that
the linear theory approximates the capillary waves of the interface by those 
of a free interface with some suitable long-wavelength cutoff, {\em i.e.}, 
they are taken to be distributed symmetrically about the mean interface 
position. The failure of the theory to describe the details of the
profiles of $\chi_z$ and $\chi_{zz}$ presumably reflects the fact that
the capillary waves are in fact asymmetric. Nevertheless, the main
features of the profiles are captured by the theory.

The centers of the peaks are slightly more distant from the surface than 
$\langle l \rangle$ in Fig. \ref{ll}, but the difference is not significant 
(up to three layers at $(H_0-H)/V=0.0007$). According to the theoretical
prediction (\ref{chil}) and (\ref{chill}), the heights of the peaks should 
diverge with $1/(H_0-H)$ with different logarithmic corrections.
Our data are shown in Figs. \ref{chin23m} and \ref{chinn23m}.
The maxima of the layer-bulk susceptibility are best fitted by the
simple $1/(H-H_0)$ behavior, which the theory predicts as long as the 
interfacial width is dominated by the intrinsic width $W_0$. 
In the regime $(H_0-H)/V<0.005$, where the capillary wave broadening
of the interface becomes significant, the data are also consistent 
with the logarithmically corrected version 
$\chi_z^{max} \propto 1/(H_0-H)\sqrt{|\ln (H_0-H)|}$
(see Fig. \ref{chin23m}). 

The analysis of the layer-layer susceptibility is more subtle. 
From a double logarithmic plot of the raw data, one is tempted to conclude
that the predicted $1/(H_0-H)$ 
behavior is not valid; the data rather
suggest a divergence with a critical exponent $0.63$ 
(Fig. \ref{chinn23m}, inset). However, since we are not aware of any 
theoretical explanation which could motivate such an exponent, we believe 
that the apparent power law behavior over roughly two decades of $(H_0-H)$ 
is most likely accidental. Looking at the values of $\chi_{zz}$ close to the 
center of the slab (Fig. \ref{chiprof}b)), one recognizes that the contribution
of bulk fluctuations to $\chi_{zz}$ is significant even close to $H_0$. 
The situation is complicated by the fact that the bulk fluctuations increase 
considerably in the vicinity of $H_0$, although their amplitude does not 
diverge. Within the crude approximation that the capillary waves of the 
interface and the bulk fluctuations are uncorrelated, one can subtract 
the latter as ``background''. The thereby corrected data agree reasonably well 
with the theory, especially when taking into account the logarithmic 
correction $\chi_{zz}^{max} \propto 1/(H_0-H)|\ln(H_0-H)|$ 
(Fig. \ref{chinn23m}). 

\begin{figure}
\noindent
\hspace*{0.5cm}\fig{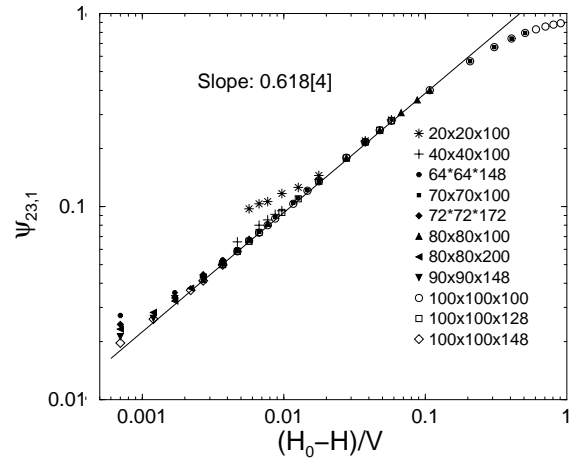}{75}{63}{ 
\vspace*{0.cm} 
\caption{
Order parameter $\psi_{23,1}$ at the surface (first layer) vs. $(H_0-H)/V$ for 
different system sizes $L\times L \times D$ as indicated. Solid line indicates
power law with the exponent $\beta_1 = 0.618$.
}
\label{psi230}
}
\end{figure}

We proceed to study the properties of the system directly at the surface.
Figure \ref{psi230} shows the order parameter $\psi_{23,1}$ in the first layer 
(recalling that the top (zeroth) 
layer is discarded) as a function of $(H_0-H)$ for various 
\noindent

\begin{figure}
\hspace*{0.cm}\fig{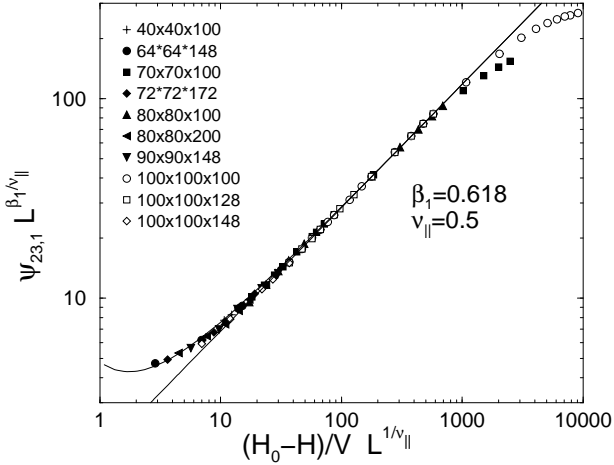}{80}{68}{
\vspace*{-0.cm} 
\caption{
Finite-size scaled plot of the surface order parameter $\psi_{23,1}$ 
vs. $(H_0-H)/V$ for system sizes $L\times L \times D$ as indicated. Data were
scaled with exponents $\nu_{\parallel}=1/2$ and $\beta_1=0.618$. Dashed
line shows the finite size scaling function predicted by eqn. (51).
}
\label{psi230s}
}
\end{figure}
\noindent
system sizes. One notices finite size effects if the dimension $L$ 
parallel to the interface is small. As long as $L$ is large enough, the data 
exhibit a power law behavior with 
the exponent $\beta_1=0.618[4]$. 
We emphasize that $\beta_1$ clearly differs from 1/2 here. It is close to the
value $\beta_1=0.64$ found by Schweika et al in their simulations of 
surface induced disorder in fcc-alloys\cite{schweika}. As discussed in
section \ref{surface}, several factors may lead to such a nonuniversal
exponent -- capillary wave fluctuations, and/or the presence of a
length scale $\lambda_J > \xi_b/2$, which competes with the correlation
length $\xi_b$ and would have to be associated with the nonordering 
composition fluctuations in the case of the symmetry preserving (110) surface.
Using eqn. (\ref{b1_3}), we can derive upper bounds for the capillary
parameter, $\omega < 0.236$, and for  $\lambda_J$, $\lambda_J/\xi_b < 0.618$. 

After applying finite size scaling with the exponents $\beta_1$ and
$\nu_{\parallel} = 1/2$ (cf. eqn. (\ref{m1s1}), the data collapse onto a 
single master curve. The form of the latter can be calculated from 
eqn. (\ref{m1s1}),
\begin{equation}
\psi_{23,1} \: L^{\beta_1/\nu_{\parallel}}
\propto \frac{x^{r/2+\omega/r}}{(x+1)^{\omega/2}} 
\; e^{2 \pi \omega /x}
\end{equation}
with $x \propto (H_0-H) L^{1/\nu_{\parallel}}$ and
$r = \mbox{max}(1, 2 \lambda_J/\xi_b)$, where the two unknown 
proportionality constants are fit parameters and $\omega=0.236$ was
used (the result is only very barely sensitive to the choice of $\omega$).
Fig. \ref{psi230s} shows that the data agree nicely with the theoretical 
prediction.

\begin{figure}
\noindent
\hspace*{0.cm}\fig{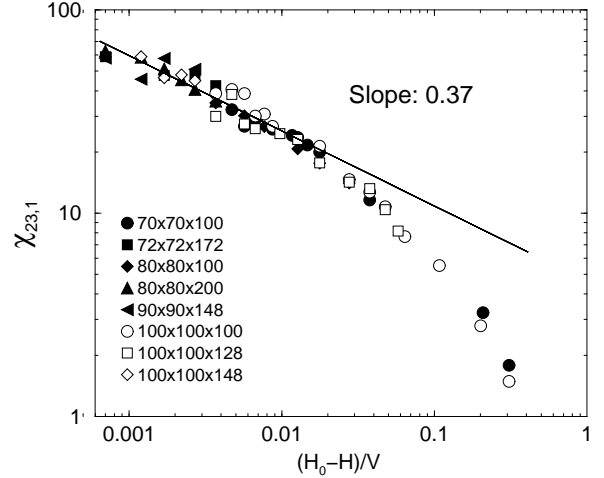}{78}{68}{
\vspace*{-0.cm} 
\caption{
Surface layer-bulk susceptibility per site of the order parameter $\psi_{23}$ 
vs. $(H_0-H)/V$ for different system sizes as indicated.
Solid line marks a power law with the exponent $\gamma_1=0.37$.
}
\label{chin230}
}
\end{figure}

Figure \ref{chin230} shows the layer-bulk susceptibility at the surface
for the order parameter $\psi_{23}$. According to eqn. (\ref{chi1}),
it should diverge with the exponent $\gamma_1 = 1-\beta_1 = 0.382$.
Indeed, the fit to our data in the region $(H_0-H)/V < 0.02$ yields 
$\gamma_1 =0.37[5]$. 
In the case of the layer-layer susceptibility,
the theory (\ref{chi11}) predicts $\gamma_{11} = 1-2\beta_1 =-0.236$, 
{\em i.e.}, $\chi_{11}$ does not diverge at the phase
transition. In fact, it first increases as $H_0$ is approached, 

\noindent
but then decreases for $(H_0-H)/V < 0.02$ (not shown). The layer-layer 
susceptibility at the surface here behaves in a 
similar way as observed by Schweika {\em et al} in their studies of surface 
induced disorder at the (111) surface of an fcc-based alloy\cite{schweika}.

\begin{figure}
\noindent
\hspace*{0.5cm}\fig{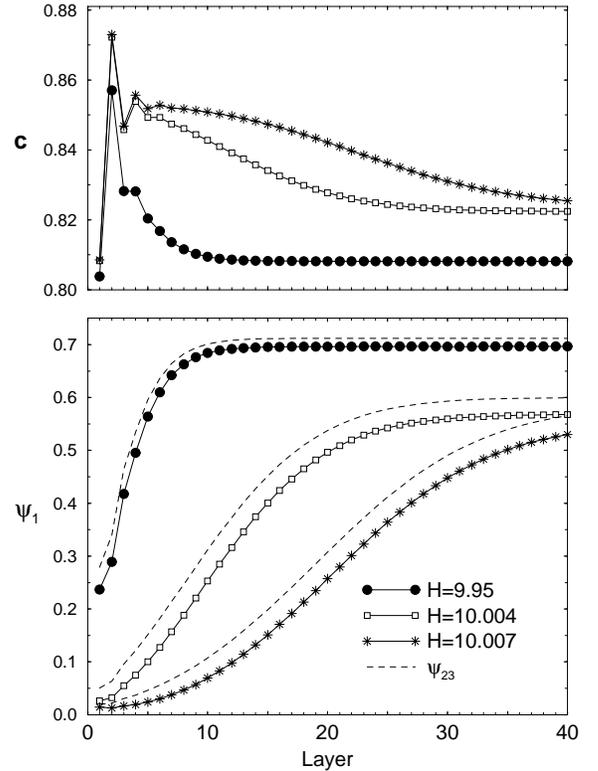}{75}{104}{
\vspace*{-0.1cm} 
\caption{
Profiles of the total composition $c = (\langle S \rangle+1)/2$ (top) and 
of the order parameter $\psi_1$ (bottom) for different fields $H$ in units
of $V$ as indicated. 
Top (zeroth) layer is not shown ($c(0)\equiv 1, \psi_1(0)\equiv(0)$).
Thin dashed lines with squares show for comparison the profiles of 
$\psi_{23}$ from Fig. 5.
}
\label{profiles}
}
\end{figure}
\noindent

\begin{figure}
\noindent
\hspace*{0.cm}\fig{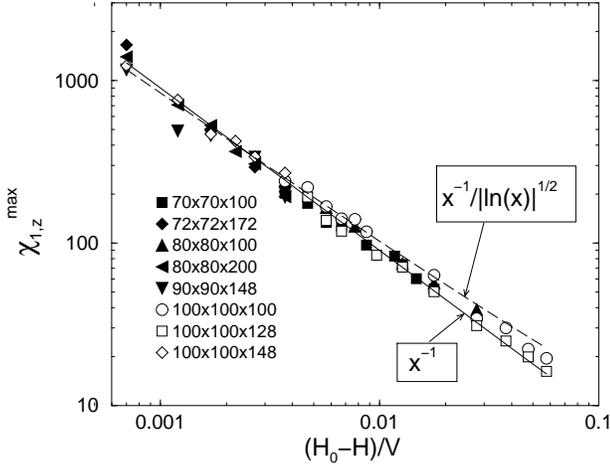}{80}{65}{ 
\vspace*{-0.cm}
\caption{
Maximum of the layer-bulk susceptibility $\chi_z$ per site of
the order parameter $\psi_{1}$ in units of $k_B T$ vs. $(H_0-H)/V$ 
for different system sizes as indicated.
Solid line shows a fit to a $(H_0-H)^{-1}$ behavior, 
and dashed line the same with the appropriate logarithmic correction.
}
\label{chin1m}
}
\end{figure}
\noindent

\begin{figure}
\noindent
\hspace*{0.cm}\fig{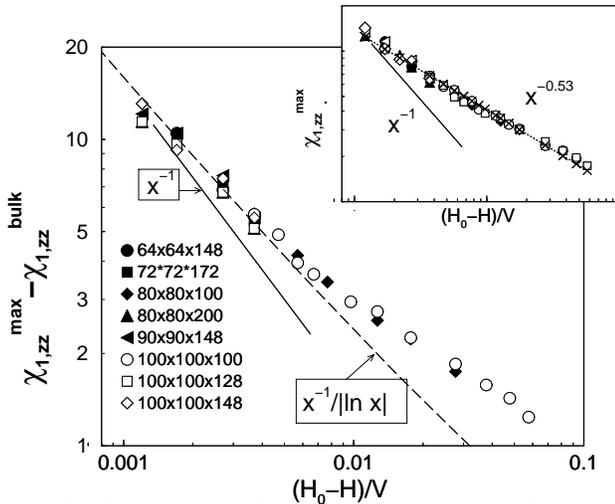}{80}{70}{
\vspace*{-0.2cm}
\caption{
Maxima of the layer-layer susceptibility $\chi_{zz}$ per site of
the order parameter $\psi_{1}$ minus bulk contribution
in units of $k_B T$ vs. $(H_0-H)/V$, for different system sizes as indicated.
Solid line indicates the slope of 
$(H_0-H)^{-1}$, and dashed line the whole theoretical prediction including
the logarithmic correction.
Inset shows bare data, with a fit to a power law behavior (dotted line).
}
\label{chinn1m}
}
\end{figure}
\noindent

\subsection{(110) Surfaces -- B2 order}

From the results discussed so far, we conclude that the behavior of the 
order parameter $\psi_{23}$ can be understood nicely within the effective 
interface theory of critical wetting. However, we shall see that this holds
only in part for the second order parameter, $\psi_1$. 

Fig. \ref{profiles} shows profiles of $\psi_1$ for different fields $H$. 
They resemble those of $\psi_{23}$, in particular the inflection point of the 
profiles is located approximately at the same distance from the surface. 
The upper part of Fig. \ref{profiles} displays profiles of the total 
concentration $c$ of $A$ particles (eqn. (\ref{cc}). They exhibit some 
characteristic, $H$-independent oscillations in the first four layers, 
and the $A$ concentration is slightly enhanced in the disordered region.
However, the overall variation is rather small.

The layer susceptibility profiles of the order parameter $\psi_1$
are qualitatively similar to those of $\psi_{23}$ and not shown here. 
Fig. \ref{chin1m} demonstrates that the maximum of the layer-bulk
susceptibility evolves with the field $H$ as theoretically predicted,
$\chi_z^{max} \propto 1/(H_0-H)\sqrt{|\ln (H_0-H)|}$. 
In the case of the layer-layer susceptibility, the agreement with the 
theoretically expected behavior 
$(\chi_{zz}^{max} - \chi_{zz}^{bulk}) \propto 1/(H_0-H)|\ln(H_0-H)|$ 
is not quite as convincing, but 
the data are still consistent with the theory 
for $(H-H_0)/V < 0.01$ (Fig. \ref{chinn1m}). 
Note that the bare values of $\chi_{zz}^{max}$  would again rather suggest 
a power law, $\chi_{zz}^{max} \propto (H_0-H)^{-0.53}$ (Fig. \ref{chinn1m},
inset), which is however most likely accidental.

\begin{figure}
\noindent
\hspace*{0.5cm}\fig{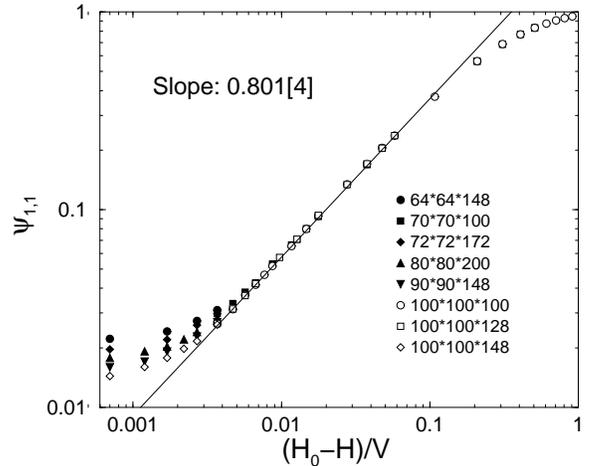}{77}{63}{ 
\vspace*{-0.1cm} 
\caption{
Order parameter $\psi_{1}$ at the surface vs. $(H_0-H)/V$ for different
system sizes $L\times L \times D$ as indicated. Solid line shows
power law with the exponent $\beta_1 = 0.801$.
}
\label{psi10}
}
\end{figure}

\begin{figure}
\noindent
\hspace*{0.cm}\fig{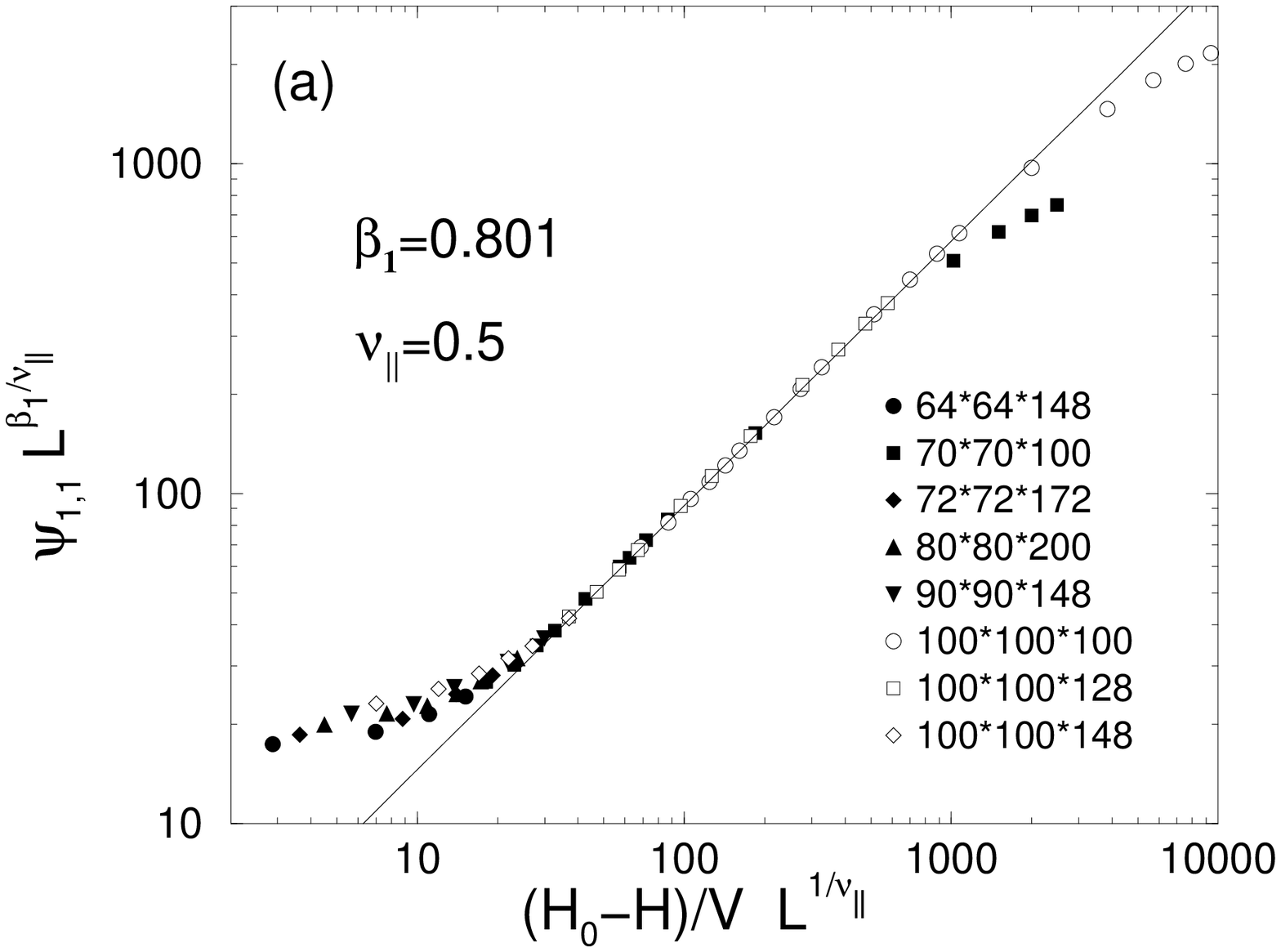}{80}{65}{}\\
\noindent
\hspace*{0.2cm}\fig{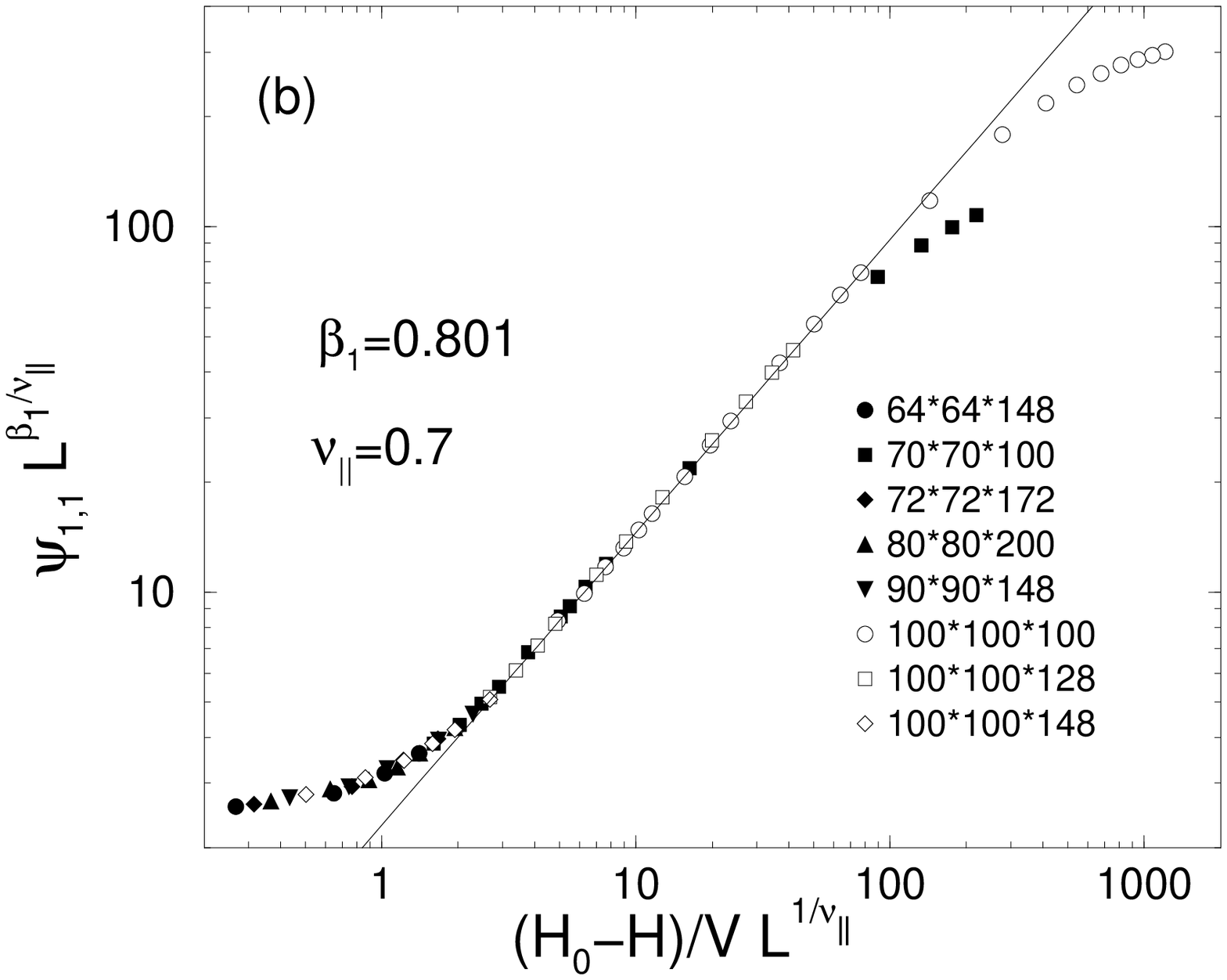}{75}{65}{
\vspace*{0.1cm} 
\caption{
Finite-size scaled plots of the order parameter $\psi_{1}$ at the surface 
vs. $(H_0-H)/V$ for system sizes $L\times L \times D$ as indicated. 
Exponents are $\beta_1=0.801$, $\nu_{\parallel}=0.5$ in (a), and
$\nu_{\parallel}=0.7$ in (b). 
}
\label{psi10s}
}
\end{figure}

Hence the behavior of the order parameter $\psi_1$ in the vicinity of the 
interface is similar to that of the order parameter $\psi_{23}$ and consistent 
with the theory of critical wetting. The agreement however does not persist
when looking right at the surface. Figs. \ref{psi10} and \ref{psi10s} 
show how the value of $\psi_1$ in the first surface layer depends on 
$(H_0-H)/V$. A power law behavior is found over one and a half decades of 
$(H_0-H)/V$, yet the exponent $\beta_1(\psi_{1})=0.801$ differs from that of 
$\psi_{23,1}$, $\beta_1(\psi_{23})=0.618$ (Fig. \ref{psi10}).
Moreover, the data for different system sizes do not collapse if
one performs finite size scaling with the exponent $\nu_{\parallel}=1/2$
(Fig. \ref{psi10s}(a)). The collapse is significantly better
if one assumes that the parallel correlation length diverges with the
exponent $\nu_{\parallel} = 0.7 \pm 0.05$ (Fig. \ref{psi10s} (b)). 

We have no explanation for these unexpected findings. The discussion in
section \ref{wetting} has shown that several surface exponents $\beta_{i,1}$ 
may be present in a system with several order parameters. Even though we
have argued that only the smallest exponent should survive in the asymptotic
limit $\mu \to 0$, the other power law contributions may conceivably 
still dominate the behavior of certain quantities over a wide range of $\mu$.
However, the critical exponent $\nu_{\parallel}$ should in all cases 
remain invariably $\nu_{\parallel}=1/2$. Our results seem to indicate that 
the behavior of the order parameter $\psi_1$ at the surface is governed 
by a length scale, which differs from that given of the interfacial 
fluctuations, but which nonetheless diverges as $H_0$ is approached.  
Note that $\nu_{\parallel} \approx 0.7$ is close to the exponent 
$\nu = 0.63$ with which the bulk correlation length diverges at an Ising type 
transition in three dimensions. Likewise, 
the exponent $\beta_1 = 0.801$
found here resembles the surface critical exponent of the ordinary transition,
$\beta_1\sim 0.8$ \cite{diehl2,KB3}. One might thus suspect that $\psi_1$ in 
the disordered surface layer becomes critical at $H_0$. However, such a co-

\begin{figure}
\noindent
\hspace*{0.5cm}\fig{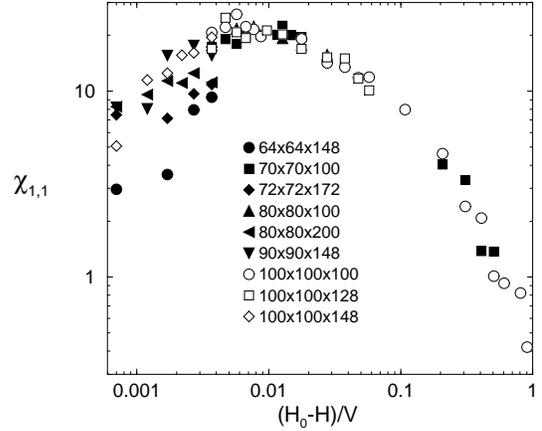}{75}{63}{
\vspace*{-0.cm} 
\caption{
Surface layer-bulk susceptibility per site of the order parameter $\psi_{1}$ 
vs. $(H_0-H)/V$ for different system sizes as indicated.
}
\label{chin0}
}
\end{figure}
\noindent
incidence would seem rather surprising.
Furthermore, we have noted earlier that 
the combination $\psi_2 \psi_3$ acts as an ordering field on $\psi_1$, 
hence $\psi_1$ cannot become critical as long as $\psi_{23}$ is not 
strictly zero.

Figure \ref{chin0} shows the layer-bulk susceptibility at the surface as
a function of $(H_0-H)/V$. It decreases as $H_0$ is approached, hence the 
scaling relation $\beta_1 + \gamma_1 = 1$ is obviously not met for the order 
parameter $\psi_{1,1}$. 

\subsection{(100) Surfaces}
\label{100}

Finally, we turn to the discussion of (100) surfaces. As already mentioned
earlier, (100) surfaces break the symmetry with respect to the order
parameter $\psi_1$, an ordering surface field coupling to this order 
parameter is allowed and thus usually present\cite{ich1,diehl1}. This field 
is often closely related to surface segregation\cite{ich1,ich2}. 
In our case, the excess component $A$ of the $DO{}_3$ segregates in the
surface layer and induces a staggered concentration field in the layers 
underneath, which is equivalent to $\psi_1$ ordering.

\begin{figure}
\noindent
\hspace*{0.5cm}\fig{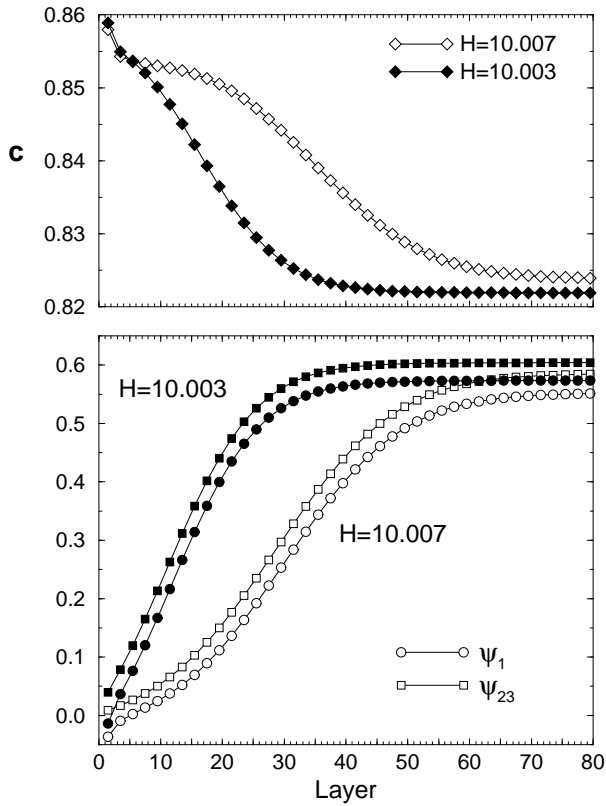}{75}{113}{
\vspace*{-0.cm} 
\caption{
Profiles of the total concentration of $A$ (top, diamonds), of the order 
parameters $\psi_1$ (bottom, circles) and $\psi_{23}$ (bottom, squares) 
at $H=10.003$ (filled symbols) and at $H=10.007$ (open symbols).
Zeroth (top) layer is not shown.
}
\label{prof_100}
}
\end{figure}

This is demonstrated in Fig. \ref{prof_100}. The order parameters and the
composition $c$ are defined based on the sublattice occupancies on two 
subsequent layers of distance $a_0/2$, starting from the first layer underneath
the surface. The top layer is again disregarded, since it is entirely
filled with $A$ or $S\equiv 1$. The profiles of $\psi_1$ clearly
display the signature of an additional ordering tendency at the surface,
which in fact reverses the sign of $\psi_1$ in the top layers.
However, the effect is rather weak and does not influence the system
significantly deeper in the bulk. The profiles can be analyzed like those
at the (110) surface, and mean interface positions and mean interfacial widths 
can be extracted to yield figures very similar to Figs. \ref{ll} and \ref{w2}.
The amplitudes of the logarithmic divergences can again be used
to estimate the bulk correlation length $\xi_b$. From the mean 
interface position, one 

\noindent

\begin{figure}
\noindent
\hspace*{0.5cm}\fig{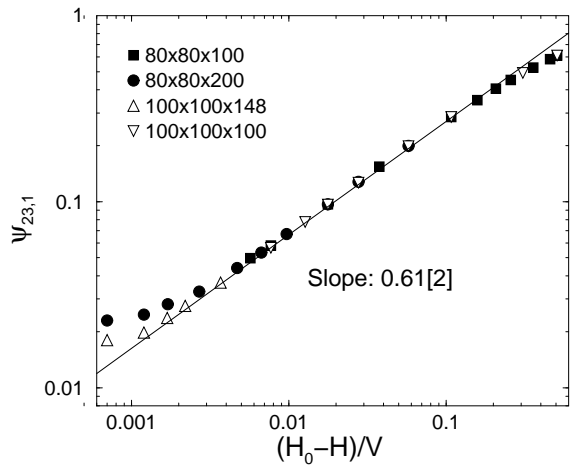}{75}{65}{ 
\vspace*{-0.cm} 
\caption{
Order parameter $\psi_{23}$ at the surface vs. $(H_0-H)/V$ for different
system sizes $L\times L \times D$ as indicated. Solid line shows
power law with the exponent $\beta_1 = 0.61$.
}
\label{psi230_100}
}
\end{figure}\noindent
calculates $4.9[7] < \xi_b/a_0 < 5.8[8]$, and from the interfacial width,
$\xi_b/a_0 > 7.5[9]$, in agreement with the values obtained for the (110) 
surface.  Likewise, the study of the layer susceptibilities at the interface 
does not offer new surprises. The maxima of the layer-bulk susceptibilities
for both $\psi_1$ and $\psi_{23}$ grow according to a power law 
$\chi_z \propto (H_0-H)^{-1}$. The layer-layer susceptibility in the
interfacial region seems to grow with a different exponent ($\sim 0.6$ like
in the case of the (110) surface), yet after subtracting the ``background'' 
the data are also consistent with the theoretically expected behavior.
Last, we study how the surface value of the order parameter $\psi_{23}$ 
evolves as the transition $H_0$ is approached. Fig. \ref{psi230_100}
shows that it vanishes according to a power law with the exponent 
$\beta_1=0.61[2]$, which is within the error the same exponent as in the case
of the (110) surface. As far as the surface behavior of $\psi_{23}$ is 
concerned, the (100) and the (110) surface are thus basically equivalent. 
The weak ordering tendency of $\psi_1$ has an at most slightly perturbing
effect on the profiles of $\psi_{23}$.

\section{Summary and Outlook}
\label{summary}

We have presented an extensive Monte Carlo study of
surface induced disorder in a simple spin lattice model for bcc-based 
binary alloys. Our work complements earlier Monte Carlo simulations of
Schweika {\em et al}\cite{schweika}, who have studied surface induced disorder 
in fcc-based alloys within a similar model.
Like these authors, we observe critical wetting behavior
with nonuniversal exponents. We have discussed our results in terms of
an effective interface model designed to describe a system with several
order parameters. In such a complex material, nonuniversal exponents may
result both from fluctuation effects and from a competition of length scales.

Due to the complicated order parameter structure in our system, however,
our data could not fully be explained within a theory which traces 
everything back to the properties of a single interface between a
disordered and an ordered phase. The theory provides a satisfactory
picture for the behavior of the order parameter describing 
the DO${}_3$ ordering, $\psi_{23}$, and in general for the structure
in the interfacial region. However, it fails to predict the behavior of the 
order parameter of B2 ordering, $\psi_1$, directly at the surface. 
Our data thus indicate that the fluctuations of $\psi_1$ at the surface 
require special treatment. Parry and coworkers\cite{swain1,parry} have 
recently suggested an approach to a theory of wetting based on an effective 
interface Hamiltonian with two ``interfaces'', the usual one separating
the phase adsorbed at the surface and the bulk phase, and a second one which 
accounts in an effective way for the fluctuations directly at the surface.
Our problem seems to call for such an approach.
Unfortunately, we are far from understanding even the constituting elements, 
the fluctuations of $\psi_1$ at the wall. We seem to observe a coupling 
between critical wetting and some kind of surface critical behavior of 
$\psi_1$, the origin of which is unclear.

Hence already our simple, highly idealized model exhibits a complex and rather
intriguing wetting behavior. In real alloys, numerous additional complications 
are present which will lead to an even richer and more interesting 
phenomenology. For example, long range 
interactions are known to influence wetting transitions significantly. 
The effect of van-der-Waals forces on wetting has been investigated in
detail\cite{wetting}. Van-der-Waals forces are important in liquid-vapour 
systems or binary fluids, but presumably irrelevant in alloys.
Instead, elastic interactions caused by lattice distortions presumably 
play an important role. 

Furthermore, real surfaces are never ideally smooth,
but have steps and islands. We have seen that the orientation of
the surface affects the surface ordering. In our study, we did not
observe dramatic differences between the (110) surface and the (100) surface. 
Nevertheless, we expect that the influence of the surface orientation on the 
wetting behavior can be quite substantial, {\em e.g.}, in situations with 
strong surface segregation, or if surface orientations are involved 
which also break the symmetry with respect to the DO${}_3$ order 
({\em e.g.}, the (111) surface).
Likewise, we can expect that steps and islands will affect the ordering
and the wetting properties of the alloy. It is well known in general that 
the wetting behavior on corrugated or rough surfaces differs from
that on smooth surfaces\cite{borgs,netz,swain2}. In addition, even a few steps or
islands on an otherwise smooth, but symmetry breaking surface of an alloy 
can have a dramatic effect on the ordering behavior, since every step
changes the sign of the ordering surface field.

\section*{Acknowledgments}

We wish to thank M. M\"uller and A. Werner for helpful discussions.
F.F.H. acknowledges financial support from the Graduiertenf\"orderung of 
the Land Rheinland Pfalz, and F.S. has been supported from the Deutsche 
Forschungsgemeinschaft through the Heisenberg program.



\begin{thebibliography}{99}
\bibitem{fn1} Under certain, rather unusual circumstances, the
  critical wetting point could be approached on a complete wetting path,
  however (see section \ref{scaling}).
\bibitem{fn2} Our multispin code differs from the one described in \cite{bhanot}
  in that we could not make use of bit operations, unfortunately, due to the 
  complicated nature of the interactions in our model.
\end{thebibliography}
\end{document}